\documentclass[apj]{emulateapj}
\usepackage{apjfonts}
\usepackage{float}
\usepackage{booktabs}

\shorttitle{}
\shortauthors{Nelson et al.}

\begin{document}

\gdef\ha{H$\alpha$}
\gdef\ew{${\rm EW}({\rm H}\alpha)$}
\gdef\msun{M$_{\odot}$}
\gdef\kms{km/s}
\gdef\f{$H_{F140W}$}
\gdef\m{M$_*$}

\title{Where stars form: inside-out growth and coherent star formation from
HST \ha\, maps of 2676 galaxies across the main sequence at $z\sim1$}

\author{Erica June Nelson\altaffilmark{1}, 
Pieter G.\ van Dokkum\altaffilmark{1}, 
Natascha M. F\"orster Schreiber\altaffilmark{2},
Marijn Franx\altaffilmark{3},
Gabriel B. Brammer\altaffilmark{4},
Ivelina G. Momcheva\altaffilmark{1},
Stijn Wuyts\altaffilmark{2},
Katherine E. Whitaker\altaffilmark{5},
Rosalind E. Skelton\altaffilmark{6},
Mattia Fumagalli\altaffilmark{3},
Mariska Kriek\altaffilmark{7},
Ivo Labb\'e\altaffilmark{3},
Joel Leja\altaffilmark{1},
Hans-Walter Rix\altaffilmark{8},
Linda J. Tacconi\altaffilmark{2},
Arjen van der Wel\altaffilmark{8},
Frank C. van den Bosch\altaffilmark{1},
Pascal A. Oesch\altaffilmark{1},
Claire Dickey\altaffilmark{1},
Johannes Ulf Lange\altaffilmark{1}
}

\altaffiltext{1}{Astronomy Department, Yale University, 
New Haven, CT 06511, USA}
\altaffiltext{2}{Max-Planck-Institut f\"ur extraterrestrische Physik,
Giessenbachstrasse, D-85748 Garching, Germany}
\altaffiltext{3}{Leiden Observatory, Leiden University, Leiden, The
Netherlands}
\altaffiltext{4}{Space Telescope Science Institute, 3700 San Martin Drive,
Baltimore, MD 21218, USA}
\altaffiltext{5}{Astrophysics Science Division, Goddard Space Flight Center,
Greenbelt, MD 20771, USA}
\altaffiltext{6}{South African Astronomical Observatory, P.O.\ Box 9,
Observatory, 7935, South Africa}
\altaffiltext{7}{Department of Astronomy, University of California,
Berkeley, CA 94720, USA}
\altaffiltext{8}{Max Planck Institute for Astronomy (MPIA), K\"onigstuhl 17,
69117, Heidelberg, Germany}

\begin{abstract}
We present \ha\ maps at 1\,kpc spatial resolution for star-forming galaxies 
at $z\sim1$, made possible by the WFC3 grism on HST.
Employing this capability over all five 3D-HST/CANDELS fields provides 
a sample of $2676$ galaxies enabling a division into subsamples 
based on stellar mass and star formation rate. 
By creating deep stacked \ha\
images, we reach surface brightness limits of
$1\times10^{-18}\,\textrm{erg}\,\textrm{s}^{-1}\,\textrm{cm}^{-2}\,\textrm{arcsec}^{-2}$,
allowing us to map the distribution of ionized gas out to greater than 10\,kpc for 
typical L$^*$ galaxies at this epoch.
We find that the spatial extent of the \ha\ distribution
increases with stellar 
mass as $r_{{\rm H}\alpha}=1.5(M_*/10^{10}M_{\odot})^{0.23}$ kpc. 
Furthermore, the \ha\ emission is more extended than the stellar
continuum emission, consistent with inside-out assembly of
galactic disks.
This effect, however, is mass dependent with $r_{{\rm H}\alpha}/r_{*}
=1.1 (M_*/10^{10}M_{\odot})^{0.054}$,
such that at low masses $r_{{\rm H}\alpha}\sim r_{*}$. 
We map the \ha\ distribution as a function of SFR(IR+UV) and 
find evidence for
`coherent star formation' across the SFR-\m\, plane: 
above the main sequence, \ha\ is enhanced at all
radii; below the main sequence, \ha\ is depressed at all radii. 
This suggests that at all masses the physical processes driving the
enhancement
or suppression of star formation act throughout the disks of galaxies. 
It also confirms that the scatter in the 
star forming main sequence is real and caused by variations in the star 
formation rate at fixed mass. 
At high masses ($10^{10.5}<M_*/M_{\odot}
<10^{11}$), above the main sequence,
\ha\ is particularly enhanced in the center, indicating that gas is being funneled  
to the central regions of these galaxies to build bulges and/or supermassive black holes.
Below the main sequence, the star forming disks are more compact and 
a strong central dip in the EW(${\rm H}\alpha$), and the inferred
specific star formation rate, appears. 
Importantly though, across the entirety of 
the SFR-\m\, plane we probe, the absolute star formation rate as
traced by H$\alpha$ is always 
centrally peaked, even in galaxies below the main sequence. 
 
\end{abstract}

.


\section{Introduction}

The structural formation history of galaxies
is written by the spatial distribution of their star formation through cosmic time. 
Recently, the combination of empirical modeling and observations of 
the scaling relation between stellar mass and star formation rate has enabled 
us to constrain the build up of stellar mass in galaxies over a large fraction
of cosmic time ({Yang} {et~al.} 2012; {Leja} {et~al.} 2013; {Behroozi} {et~al.} 2013; {Moster}, {Naab}, \& {White} 2013; {Lu} {et~al.} 2014; {Whitaker} {et~al.} 2014).
The dawn of Wide Field Camera 3 (WFC3) on 
the Hubble Space Telescope (HST)
has enabled us to map the structural growth
of this stellar mass content of galaxies at high fidelity over a large fraction
of the history of the universe (e.g. {Wuyts} {et~al.} 2011a, 2012; {van der Wel} {et~al.} 2012; van~der Wel {et~al.} 2014a, 2014b; Bruce {et~al.} 2014; Boada {et~al.} 2015; Peth {et~al.} 2015).
It has become clear that the physical sizes of galaxies
increase with cosmic time as the universe 
expands (Giavalisco, Steidel, \&  Macchetto 1996; Ferguson {et~al.} 2004; Oesch {et~al.} 2010; Mosleh {et~al.} 2012; Trujillo {et~al.} 2006; {Franx} {et~al.} 2008; {Williams} {et~al.} 2010; {Toft} {et~al.} 2007; Buitrago {et~al.} 2008; {Kriek} {et~al.} 2009; van~der Wel {et~al.} 2014a).
For star forming galaxies, with increasing stellar mass, the disk scale length 
increases as does the prominence of the bulge (e.g. {Shen} {et~al.} 2003; {Lang} {et~al.} 2014).
The picture that has emerged from these studies is that most galaxies 
form their stars in disks growing inside out 
({Wuyts} {et~al.} 2011a, 2013; van~der Wel {et~al.} 2014b; {Abramson} {et~al.} 2014).

In the canonical paradigm, inside-out growth is a consequence of 
the dark mater halo properties of the galaxies.
Galaxies are thought to accrete their gas from the cosmic web at a 
rate throttled by the mass of their dark matter halo 
(e.g. White \& Rees 1978; {Dekel} {et~al.} 2013). 
The gas cools onto the disk of the galaxy and forms stars with 
a radial distribution set by the angular momentum distribution 
of the halo ({Fall} \& {Efstathiou} 1980; Dalcanton, Spergel, \&  Summers 1997; van~den Bosch 2001).
As the scale factor of the universe increases, so does the spatial extent 
of the gas (Mo, Mao, \& White 1998); galaxies were smaller in the past and grow larger with 
time, building up from the inside-out.
However, the actual formation of galaxies in a cosmological 
context is more complex (e.g., van~den Bosch 2001; {Hummels} \& {Bryan} 2012).  
Recently, significant progress has been made by the creation of realistic
disk galaxies in hydrodynamical simulations
({Governato} {et~al.} 2010; Agertz, Teyssier, \& Moore 2011; Guedes {et~al.} 2011; {Brooks} {et~al.} 2011; {Stinson} {et~al.} 2013; {Aumer} {et~al.} 2013; Marinacci, Pakmor, \&  Springel 2013) 
and combining theory and observations in a self-consistent framework
(Keres {et~al.} 2009; Dekel \& Birnboim 2006; {Dekel} {et~al.} 2009b; {Genzel} {et~al.} 2008, 2011; {F{\"o}rster Schreiber} {et~al.} 2009, 2011a; {Wuyts} {et~al.} 2011b, 2011a).
How gas is accreted on to galaxies (e.g.  Brooks {et~al.} 2009; {Sales} {et~al.} 2012) 
and feedback (e.g.  Keres {et~al.} 2005; {Sales} {et~al.} 2010; {{\"U}bler} {et~al.} 2014; {Nelson} {et~al.} 2015; {Genel} {et~al.} 2015) 
have been shown to be essential ingredients.
However, precisely what physical processes drive the sizes, 
morphologies, and evolution of disk galaxies is still a matter 
of much debate (see, e.g., {Dutton} \& {van den Bosch} 2012; Scannapieco {et~al.} 2012).

Furthermore, the evidence for this picture is indirect: we do not actually observe
star formation building up different parts of these galaxies. Instead, we infer it 
based on empirically linking galaxies across cosmic time and tracking 
radial changes in stellar surface densities and structural parameters
({van Dokkum} {et~al.} 2010; {Wuyts} {et~al.} 2011a; van Dokkum {et~al.} 2013; Patel {et~al.} 2013; van~der Wel {et~al.} 2014a; Brennan {et~al.} 2015; Papovich {et~al.} 2015).  
However, this method has uncertainties due to scatter in stellar mass 
growth rates and merging (e.g. {Leja} {et~al.} 2013; {Behroozi} {et~al.} 2013). 
Furthermore, migration and secular evolution may have changed the 
orbits of stars after their formation such that they no longer live in 
their birthplaces (e.g., Ro{\v s}kar {et~al.} 2008).

The missing piece is a direct 
measurement of the spatial distribution of star formation within galaxies. 	
This is crucial to understanding the integrated relations of galaxy growth 
between SFR and \m.
The spatial distribution of star formation yields insights
into what processes drive the star formation activity, evolution of 
stellar mass, and the relation between them. It helps to disentangle 
the role of gas accretion, mergers, and secular evolution on the 
assembly history of galaxies. 
Furthermore, this provides a test of inside-out growth which 
appears to be a crucial feature of galaxy assembly history. 

What is required is high spatial resolution maps of star formation and stellar continuum 
emission for large samples of galaxies while they were actively forming 
their disks.
The \ha\, flux scales with the quantity of 
ionizing photons produced by hot young stars, serving as an excellent 
probe of the sites of ongoing star formation activity ({Kennicutt} 1998). 
A number of large surveys have used \ha\ to probe the growth 
of evolving galaxies, including recently: 
HiZELS ({Geach} {et~al.} 2008; {Sobral} {et~al.} 2009),
WISP ({Atek} {et~al.} 2010), 
MASSIV ({Contini} {et~al.} 2012), 
SINS/zC-SINF ({F{\"o}rster Schreiber} {et~al.} 2006, 2009),
KROSS, {Stott} {et~al.} (2014), and 
KMOS$^{3D}$ ({Wisnioski} {et~al.} 2015).
Broadband rest-frame optical imaging provides 
information on the stellar component. 
The spatial distribution of this stellar light contains
a record of past dynamical processes and the 
history of star formation. The comparison of the spatial distribution of 
ionized gas and stellar continuum emission thus provides an essential 
lever arm for constraining the structural assembly of galaxies. 
This potent combination shed light on the turbulent early phase of 
massive galaxy growth at $z\sim2$ 
({F{\"o}rster Schreiber} {et~al.} 2011a; {Genzel} {et~al.} 2014a; {Tacchella} {et~al.} 2015b, 2015a),
and the spatially-resolved star-forming sequence ({Wuyts} {et~al.} 2013). 
To apply this same methodology to a global structural analysis requires 
high spatial resolution spectroscopic measurements for a large sample of 
galaxies. An ideal dataset would also contain broadband optical imaging 
with the same high spatial resolution to allow for robust comparison of 
the spatial distribution of ionized gas and stellar continuum emission. 

This has now become possible with the WFC3 
grism capability on HST. 
The combination of WFC3's high spatial resolution and 
the grism's low spectral resolution provides spatially resolved 
spectroscopy. Because this spectrograph is slitless, 
it provides a spectrum for every object in its field of view. 
This means that for every object its field of view and wavelength coverage, 
the grism can be used to create a high spatial resolution
emission line map.
The 3D-HST legacy program utilizes this powerful feature for a 248 orbit 
NIR imaging and grism spectroscopic survey over the five CANDELS fields
({van Dokkum} {et~al.} 2011; {Brammer} {et~al.} 2012a, Momcheva et al. in prep).
In this paper, we use data from the 3D-HST survey to map the spatial 
distribution of \ha\ emission (a tracer of star formation)
and \f\ stellar continuum emission 
(rest-frame 7000\AA, a proxy for the stellar mass)
for a sample of 2676 galaxies at 0.7<z<1.5. 
The \ha\, and stellar continuum are resolved on scales of 0.13". 
This represents the largest survey to date
of the spatially resolved properties of the \ha\ distribution in galaxies at any epoch. 
This spatial resolution, corresponding to $\sim1$\,kpc, is 
necessary for structural analysis and only possible from the ground with 
adaptive optics assisted observations on 10m class telescopes. 
This dataset hence provides a link between the high spatial resolution 
imaging datasets of large samples of galaxies with HST and 
high spatial resolution emission line maps of necessarily small samples
with AO on large ground-based telescopes.
This study complements the large MOSDEF ({Kriek} {et~al.} 2015)
and KMOS$^{3D}$ ({Wisnioski} {et~al.} 2015)
spectroscopic surveys by providing higher spatial resolution emission line 
measurements.

We present the average surface brightness profiles of \ha\, 
and stellar continuum 
emission in galaxies during the epoch $0.7<z<1.5$. We analyze \ha\, maps for 
2676 galaxies from the 3D-HST survey to trace
the spatial distribution of star formation. Our sample cuts  
a large swath through the SFR-\m\, plane covering
two orders of magnitude in stellar mass 
$10^9<\textrm{M}_*<10^{11}$ and star formation rate 
$1<SFR<400\,\textrm{M}_\odot/$yr and encompassing the star forming 
``main sequence'' (MS).
{Wuyts} {et~al.} (2012) showed that the bright, visually striking clumps of star formation 
which appear to be common in high redshift galaxies are short-lived and 
contribute little to the integrated SFR of a galaxy. Here, we average 
over these short-lived clumps by stacking \ha\, maps. 
Stacking thousands of HST orbits provides
deep average \ha\, images that allow us to trace the \ha\, distribution 
down to a surface brightness limit of 
$1\times10^{-18}\,\textrm{erg}\,\textrm{s}^{-1}\,\textrm{cm}^{-2}\,\textrm{arcsec}^{-2}$
in our deepest stacks,
an order of magnitude fainter than previous studies in the high 
redshift universe. This enables us to measure the star formation 
surface density down to a limit of  
is $4\times10^{-4}\,\textrm{M}_\odot\,\textrm{yr}^{-1}\,\textrm{kpc}^{-2}$.
With these deep stacked images, the primary goals of this study are to derive
the average surface brightness profile and effective radius of \ha\, as 
a function of mass and star formation rate to provide insight into 
where star formation occurs in galaxies at this epoch. 

 \begin{figure*}[hbtf]
\includegraphics[width=\textwidth]{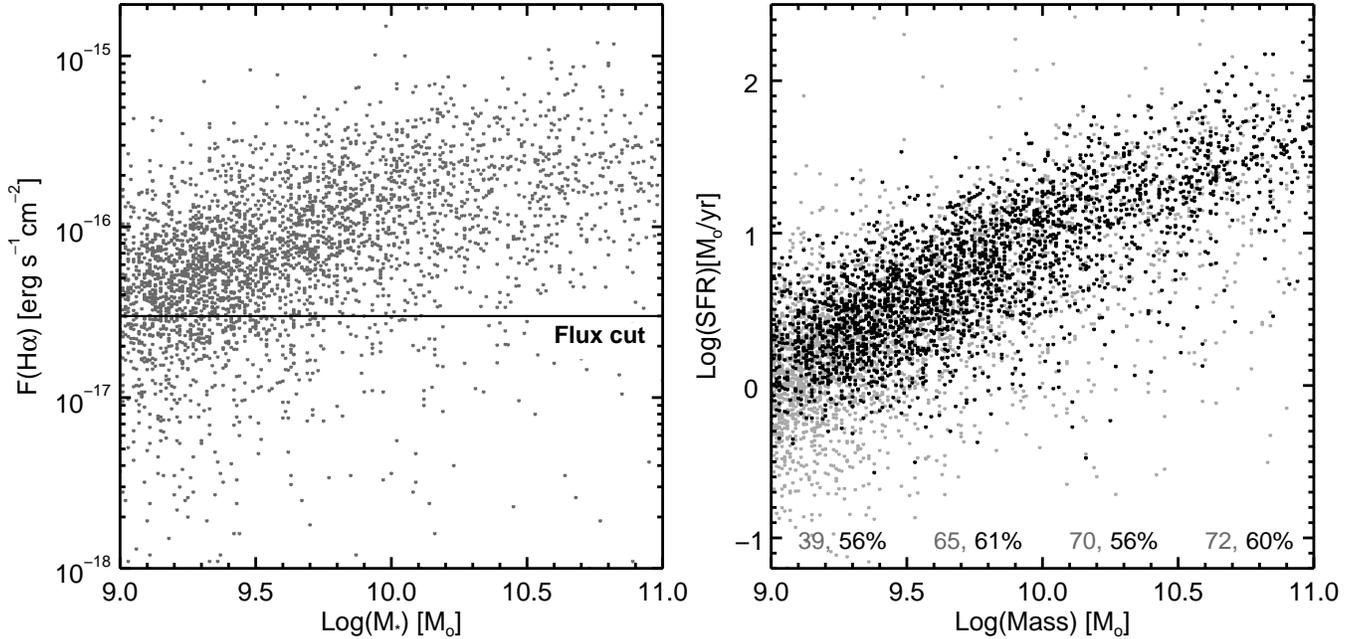}
\caption{Sample selection. 
The left panel shows the location of our flux cut with respect to the 
locus of galaxies in the F(\ha)-\m\ plane. 
The right panel shows the distribution of our sample in 
the SFR-\m\ plane. The SFRs come from the UV+IR. 
The parent sample is shown in gray; selected galaxies are shown in black.
The fraction of the total parent sample
above the \ha\, flux limit and extraction magnitude limit
are listed at the bottom in gray. 
As expected, we are significantly less complete 
at low masses and star formation rates. 
Further, about one third of galaxies are removed due to contamination 
of their spectra by other sources in the field. 
Of the galaxies above the flux and extraction limits, the
fraction remaining as part of the final selection are listed in black. 
Our sample contains 2676 galaxies from 0.7<z<1.5 spanning two 
decades in \m\, and SFR.
\label{fig:sample}}
\end{figure*}

\vspace{2.5cm}
\section{Data}

\subsection{The 3D-HST Survey}

We investigate the spatial distribution of star formation in galaxies 
during the epoch spanning $0.7<z<1.5$ across the 
$SFR-M_*$ plane using data from the 3D-HST survey. 
3D-HST is a 248 orbit extragalactic 
treasury program with HST
furnishing NIR imaging and grism spectroscopy
across a wide field ({van Dokkum} {et~al.} 2011; {Brammer} {et~al.} 2012a, Momcheva et al. in prep).
HST's G141 grism on Wide Field Camera 3 (WFC3) provides
spatially resolved spectra of all objects in the field of view. 
The G141 grism has a wavelength range of 
$1.15\mu m < \lambda < 1.65\mu m$, covering the 
 \ha\, emission line for $0.7<z<1.5$. 
Combined with the accompanying $H_{F140W}$ imaging, 3D-HST 
enables us to derive the spatial distribution of \ha\, and 
rest-frame R-band emission with matching
1\,kpc resolution for an objectively selected sample of galaxies.  

The program covers the well-studied CANDELS fields 
({Grogin} {et~al.} 2011; {Koekemoer} {et~al.} 2011)
AEGIS, COSMOS, GOODS-S, UDS,
and also includes GOODS-N (GO-11600, PI: B. Weiner.) 
The optical and NIR imaging from CANDELS in conjunction 
with the bountiful public photometric data from $0.3-24\mu$m
provide stringent constraints on the spectral energy distributions (SEDs)
of galaxies in these fields ({Skelton} {et~al.} 2014).

\subsection{Determining z, M$_*$, SFR}
This study depends on robustly determining galaxy integrated properties,
specifically M$_*$ and SFR. Both of these quantities in turn depend on
a robust determination
of redshift and constraints on the spectral energy distributions of galaxies
across the electro-magnetic spectrum. 
To do this, the photometric data was shepherded and 
aperture photometry was performed to 
construct psf-matched, deblended, $J_{F125W}/H_{F140W}/H_{F160W}$
selected photometric catalogs (see  {Skelton} {et~al.} 2014).
These photometric catalogs form the 
scaffolding of this project upon which all the remaining data products 
rest. For this study, we rely on the rest-frame colors, stellar 
masses, and star formation rates. All of these quantities were
derived based on constraints from across the electromagnetic spectrum. 

Our redshift fitting method also utilizes 
the photometry. This is probably not strictly necessary
for the sample of \ha\, line emitting galaxies used for this study,
although it helps to confirm the redshift of galaxies with only one emission 
line detected. It is crucial,
however, for galaxies without significant emission or absorption features 
falling in the grism spectrum.  
To measure redshifts, the photometry and the two-dimensional G141 
spectrum were fit simultaneously with a modified version of the 
EAzY code ({Brammer}, {van Dokkum}, \&  {Coppi} 2008). 
After finding the best redshift, emission line strengths 
were measured for all lines that fall in the grism wavelength range
(see Momcheva et al. in prep).

Galaxy stellar masses were derived using stellar population synthesis modeling
of the photometry with the FAST code (Kriek et al. 2009).
We used the Bruzual \& Charlot (2003) 
templates with solar metallicity and a 
{Chabrier} (2003) initial mass function. 
We assumed exponentially declining star formation histories and
the {Calzetti} {et~al.} (2000) dust attenuation law (see  {Skelton} {et~al.} 2014). 
Errors in the stellar mass due to contamination of the broadband 
flux by emission lines are not expected to be significant for this study 
(see appendix in {Whitaker} {et~al.} 2014).

Galaxy star formation rates in this work were computed by summing 
unobscured (UV) plus dust absorbed and re-emitted emission (IR) from young 
stars:
\begin{equation}
\textrm{SFR}=\textrm{SFR}_{UV+IR}(M_{\odot}yr^{-1})=1.09\times10^{-10}(L_{IR}+2.2L_{UV})/L_{\odot}
\end{equation}
({Bell} {et~al.} 2005).
 $L_{UV}$ is the total UV luminosity from 1216 -- 3000\,\AA.
 It is derived by scaling the rest-frame 2800\,\AA\, 
 luminosity determined from the best-fit SED with EAzY
({Brammer} {et~al.} 2008). 
 $L_{IR}$ is the total IR luminosity from $8-1000\mu$m. It is derived 
 by scaling the MIPS 24$\mu$m flux density using a luminosity-independent 
 template that is the log average of the Dale \& Helou (2002) templates with 
 $1<\alpha<2.5$ ({Wuyts} {et~al.} 2008; {Franx} {et~al.} 2008; {Muzzin} {et~al.} 2010). 
 See {Whitaker} {et~al.} (2014) for more details.

\subsection{Sample Selection}
We consider all galaxies 1) in the redshift range 
$0.7<z<1.5$ for which the \ha\, emission line falls in the G141 grism 
wavelength coverage; 2) that have stellar masses $9.0<$ log(M$_*)<11.0$, 
a mass range over which our $H-$band selected catalogs are complete;
and 3) that are characterized as star-forming according to the UVJ-color 
criterion based on 
SED shape (Labbe {et~al.} 2005; {Wuyts} {et~al.} 2007; {Whitaker} {et~al.} 2011). 
The UVJ selection separates quiescent galaxies from star forming galaxies using
the strength of the Balmer/4000\AA\, break which is sampled by the rest-frame
$U-V$ and $V-J$ colors. 
These three criteria result in a parent sample of 8068 star-forming galaxies.  
The grism spectra are fit down to $H_{F140W}=24$, 
trimming the sample to 6612.

We select galaxies based on a quite generous cut in \ha\, Flux:
 F(\ha) $>3\times10^{-17}$erg/s/cm$^2$,
This limit corresponds to a median signal to noise $S/N(H\alpha)=2$
and sample of 4314 galaxies. 
Galaxies with lower \ha\, fluxes were removed as they may 
have larger redshift errors. 
We note here that this sample is \ha-limited, not \ha-selected. 
That is, it is a mass-selected sample of star-forming galaxies where we
require an \ha\, flux to ensure only galaxies with correct redshifts are included.
As a result of the flux and grism extraction limits, we are less complete
at low masses and star formation rates. 
We exclude 178 galaxies which were flagged as having 
bad GALFIT ({Peng} {et~al.} 2002) fits in the van~der Wel {et~al.} (2014a) catalogs,
often indicative of oddities in the photometry. 
We identify galaxies that are likely to host active galactic nuclei (AGN)
 as sources with X-ray luminosity 
$L_x>10^{42.5}{\rm erg\,\,s}^{-1}$
or \ha\, emission line widths of $\sigma>1000$\,\kms\ (see next section).
We remove these 57 galaxies from the sample as emission from 
AGN would complicate the interpretation of the measured \ha\ distributions.

Finally, of this sample, we discard 34\% of galaxies due to contamination 
of their spectra by the spectra of other nearby objects (see next section for more detail). 
The contaminating spectra are primarily bright stars and galaxies 
unrelated to the object, but it is possible that
this criterion might lead to a slight bias against denser environments. The 
fraction of galaxies removed from the sample due to contamination does 
not vary with stellar mass or star formation rate. 
The final sample contains 2676 galaxies and is shown in Fig.\,\ref{fig:sample}.

\vspace{2.5cm}
\section{Analysis}
\subsection{Morphological Information in the Spectrum}

\begin{figure}[hbtf]
\includegraphics[width=0.5\textwidth]{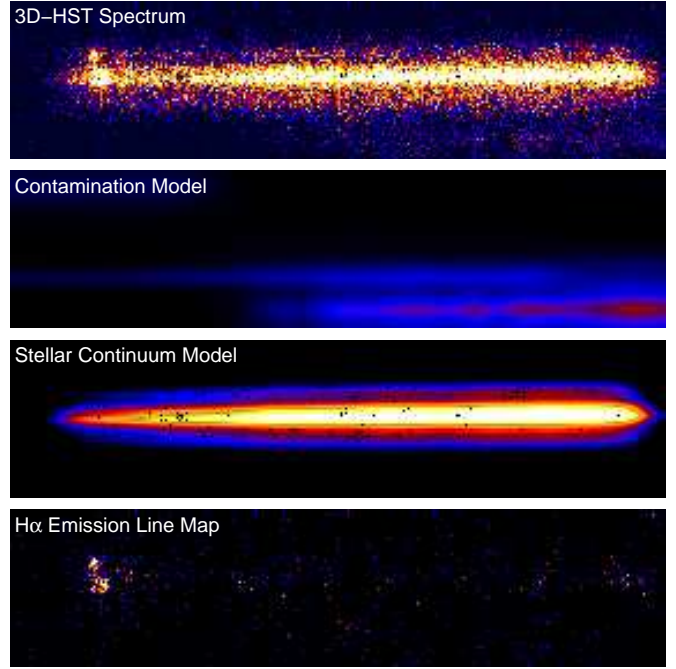}
\caption{Illustration of the creation of \ha\, emission line maps from HST WFC3
grism data. The top panel shows the 2D, interlaced grism spectrum. The second
panel shows a model for the ``contamination'':
the spectra of all objects in the field except the
object of interest. The third panel is a 2D model for the continuum emission
of the galaxy. The bottom panel is the original spectrum with the contaminating
emission from other obejcts, and the stellar continuum, subtracted. The result
is a 2D map of the line emission at the spatial resolution of HST (see Sect.\
3.2 for details).
\label{fig:makemaps}}
\end{figure}

\begin{figure*}[hbtf]
\centering 
\includegraphics[width=\textwidth]{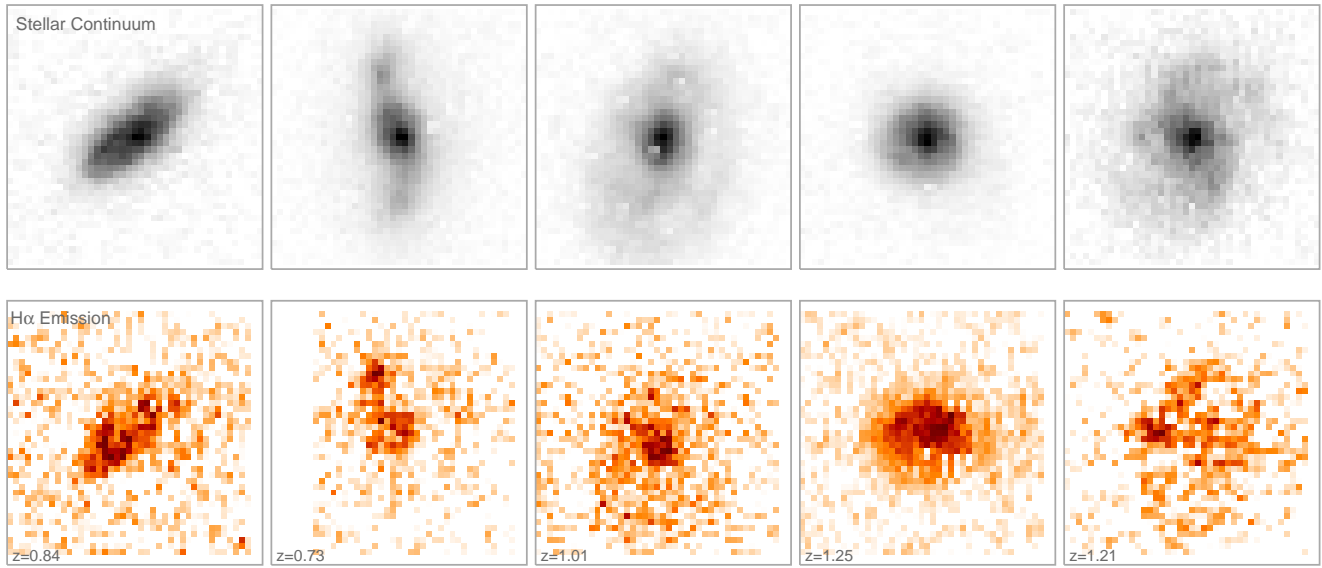}
\caption{High resolution \ha\, maps for $z\sim1$ galaxies from HST
and their corresponding rest-frame optical images. The \ha\, generally
follows the optical light but not always (see also {Wuyts} {et~al.} 2013).
\label{fig:exmaps}}
\end{figure*}

The \ha\, maps at the heart of this analysis are created
from the two-dimensional 3D-HST grism spectra. 
The creation of \ha\, emission line maps is possible as a 
consequence of a unique interaction of features:
WFC3 has high spatial resolution (0\farcs 14) and
the G141 grism has low (R$\sim$130) point source spectral resolution.
A G141 grism spectrum is a series of high resolution images of a galaxy 
taken at 46\AA\, increments and placed next to each other on the WFC3 detector. 
An emission line in such a set up effectively emerges as an image of the galaxy in that 
line superimposed on the continuum.
A resolution element for a galaxy at
$z\sim1$ corresponds to a velocity dispersion of $\sigma\sim1000$\,\kms, 
so a spectrum will only yield velocity information about 
a galaxy if the velocity difference across that galaxy is more than 1000\,\kms\,.
Few galaxies have such large line widths. Thus in general, structure in an
emission line is due to \textit{morphology}, not kinematics. 
While in a typical ground based spectroscopy, 
the shape of the emission line yields spectral information,
in our spectra it yields spatial information. 
The upshot of this property is that by subtracting the continuum from a 
spectrum, we obtain an emission line map of that galaxy.
A sample G141 spectrum is shown in Fig.\,\ref{fig:makemaps} and sample \ha\ 
maps are shown in Fig.\,\ref{fig:exmaps}.

We note that although it is generally true that the spectral axes of these \ha\, 
maps do not contain kinematic information, there is one interesting
exception: broad line AGN. 
With line widths of $>1000$\kms, the spectra of these objects do 
contain kinematic information. These sources are very easy to pick out:
they appear as point sources 
in the spatial direction and extended in the spectral direction.

Furthermore, because the WFC3 camera has no slits, 
we get a 2D spectrum of every
object in the camera's field of view. For all galaxies with 0.7<z<1.5, that have an
\ha\, emission line in G141's wavelength coverage, we obtain an \ha\, map
to the surface brightness limits. Based on our selection criteria, using this
methodology, we have a sample of 2676 galaxies at 0.7<z<1.5 
with spatially resolved \ha\, information.


\begin{figure*}[hbtf]
\includegraphics[angle=-90,width=\textwidth,trim={2cm 0 2cm 0},clip]{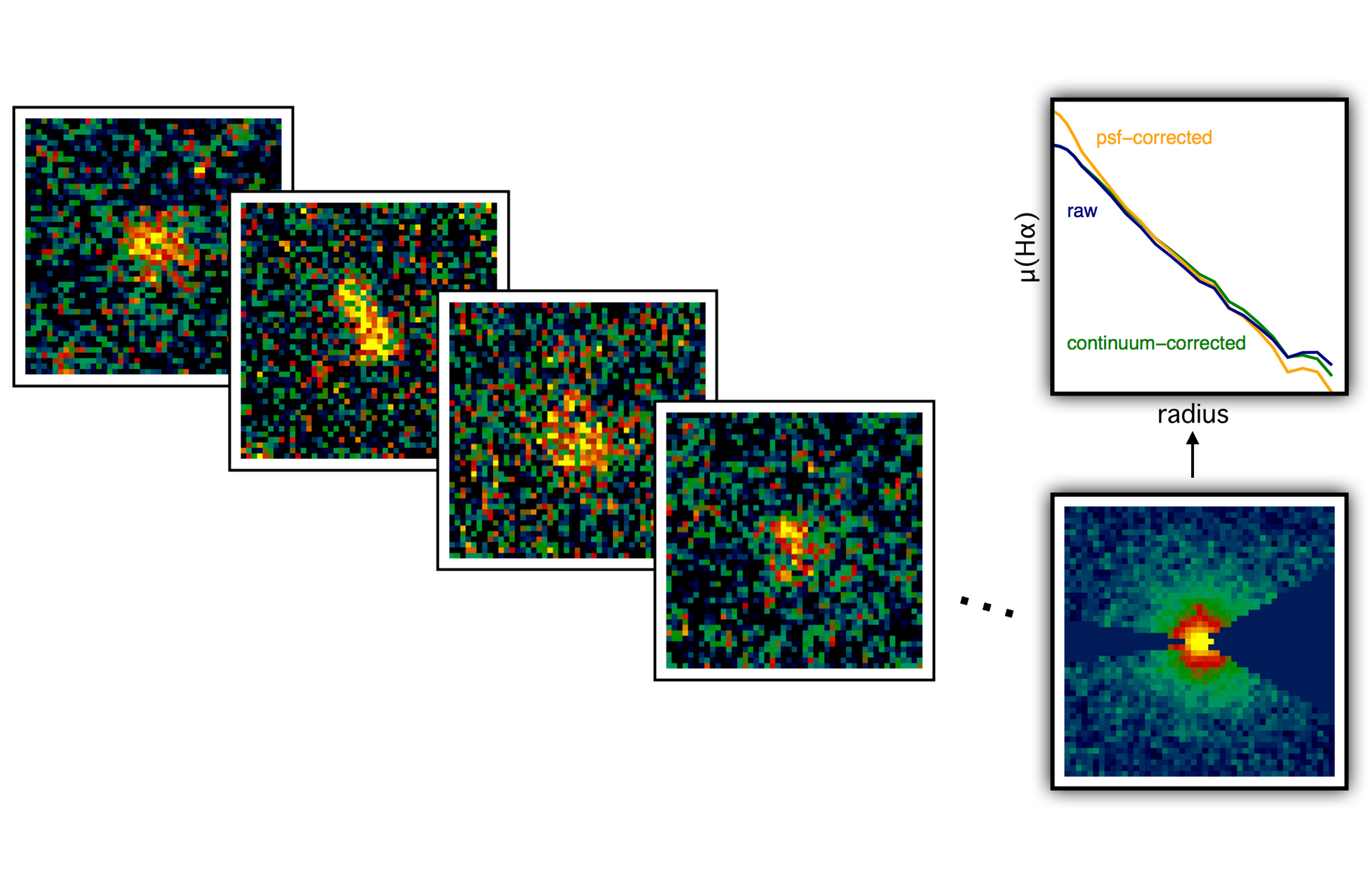}
\caption{Illustration of the creation of \ha\ image stack and the
derivation of radial profiles. The panels on the left show four of the
377 \ha\, maps that are summed to create the stack on the right. 
The stack is masked with the ``double pacman" mask 
shown, in order to mitigate the effects of redshift uncertainties and 
[S\,{\sc\,ii}] $\lambda\lambda 6716,6731$\AA\, (see \S\,3.3).
The surface brightness profiles derived from this stack are shown above it.
The raw profile is shown in black. The profile corrected for residual continuum
is shown in green and the profile corrected for the effects of the psf is shown in orange.
\label{fig:stack}}
\end{figure*}

\subsection{Making \ha\, maps}

The reduction of the 3D-HST spectroscopy with the G141 grism 
and imaging with the \f\, filter was done using a custom pipeline.
HST data is typically reduced by drizzling, but
the observing strategy of 3D-HST allows images to be interlaced instead.
With this dither pattern, four images are taken with pointing 
offsets that are multiples of half pixels.
The pixels from these four uncorrected frames are then placed on an 
output grid with 0.06" pixels ({van Dokkum} {et~al.} 2000).
Interlacing improves the preservation of 
spatial information, effectively improving the spatial resolution of the images.
Crucially, interlacing also eliminates the correlated noise caused by drizzling. 
This correlated noise is problematic for analysis of  
spectroscopic data because it can masquerade as spectral features.

Although the background levels in NIR images taken from space are 
lower than in those taken from earth, they are still significant. 
The modeling of the background in the grism data 
is complicated because it is composed 
of many faint higher order spectra. It is done using a linear combination
of three physical eigen-backgrounds: zodiacal light, metastable He emission
({Brammer} {et~al.} 2014), and scattered light from the Earth limb (Brammer et al. in prep).
Residual background structure in the wavelength direction of the 
frames is fit and subtracted along the image columns.
(For more information see  {Brammer} {et~al.} 2012a, 2014, Momcheva et al. in prep)
The 2D spectra are extracted from the interlaced G141 frames 
around a spectral trace based on a geometrical
mapping from the location of their F140W direct image positions.
A sample 2D spectrum and a pictorial depiction of the remainder of this 
subsection is shown in Fig. \ref{fig:makemaps}.

The advantage of slitless spectroscopy is also its greatest challenge:
flux from neighboring
objects with overlapping traces can contaminate the spectrum of an object 
with flux that does not belong to it. 
We forward-model contamination with a flat spectrum
based on the direct image positions and morphologies of contaminating objects.
A second iteration is done to improve the models of bright ($H<22$)
sources using their extracted spectra.
An example of this contamination model is shown in the second 
panel of Fig.\,\ref{fig:makemaps}
(See  {Brammer} {et~al.} 2012a, 2012b, 2013, Momcheva et al. in prep).
To remove contamination from the spectra, we subtract these models for 
all galaxies in the vicinity of the object of interest. 
Furthermore, for the present analysis, all regions predicted to 
have contamination which is greater than 
a third of the average G141 background value were masked. 
This aggressive masking strategy was used to reduce the uncertainty 
in the interpretation of the \ha\, maps at large radii where uncertainties in 
the contamination model could introduce systematics. 

The continuum of a galaxy is modeled by convolving the best fit SED without
emission lines with its combined $J_{F125W}/H_{F140W}/H_{F160W}$ image. 
The continuum model for our example galaxy is shown in the third panel of 
Fig.\,\ref{fig:makemaps}.
This continuum model is subtracted 
from the 2D grism spectrum, removing the continuum emission and
simultaneously correcting the emission line maps for stellar absorption. 
What remains for galaxies with $0.7<z<1.5$ is a map of their \ha\, emission. 
Five sample \ha\, maps and their corresponding \f\, images are shown 
in Fig. \ref{fig:exmaps}. 
Crucially, the \ha\, and stellar continuum images were taken 
with the same camera under the same conditions. 
This means that differences in their spatial distributions are intrinsic,
not due to differences in the PSF.
The spatial resolution is $\sim$1\,kpc for 
both the \f\, stellar continuum and \ha\, emission line maps.

The final postage stamps we use in this analysis are $80\times80$ pixels.
An HST pixel is 0.06", so this corresponds to $4.8\times4.8$" or 
$38\times38$\,kpc at $z\sim1$. Many of these postage stamps have
a small residual positive background (smaller than the noise).
To correct for this background, we compute the median of all unmasked
pixels in the 2\,kpc edges of each stamp and subtract it. This means that 
we can reliably trace the surface brightness out to 17\,kpc. 
Beyond this point, the surface brightness is definitionally zero.

\subsection{Stacking}
To measure the average spatial distribution of \ha\, during 
this epoch from $z=1.5-0.7$, we create mean \ha\, images by stacking
the \ha\, maps of individual galaxies with similar \m\, and/or SFR 
(See \S\,4\,\&\,5). 
Many studies first use \ha\ images of individual galaxies to 
measure the spatial distribution of star formation 
then describe average trends in this distribution as a function of \m\ or SFR
(e.g., {F{\"o}rster Schreiber} {et~al.} 2006; Epinat {et~al.} 2009; {F{\"o}rster Schreiber} {et~al.} 2009; {Genzel} {et~al.} 2011; {Nelson} {et~al.} 2012; {Epinat} {et~al.} 2012; {Contini} {et~al.} 2012; {Wuyts} {et~al.} 2013; {Genzel} {et~al.} 2014a).
Instead, we first create average \ha\ images by stacking galaxies 
as a function \m\ and SFR then measure the spatial distribution of 
star formation to describe trends. 
This stacking strategy leverages the
strengths of our data: \ha\, maps taken under uniform observing conditions
for a large and objectively defined sample of galaxies.
From a practical standpoint, the methodology has the advantage 
that we do not need data with very high signal-to-noise.
As a consequence, we can explore 
relatively uncharted regions of parameter space.
In particular, we can measure the radial distribution of star formation 
in galaxies across a vast expanse of the SFR-\m\ plane down to 
low masses and star formation rates. 
Additionally, we can probe the distribution of ionized gas in the
outer regions of galaxies where star formation surface densities 
are thought to be very low.

We created the stacked images by summing normalized, 
masked images of galaxies in \f\ and \ha. 
To best control for the various systematics described in 
the remainder of this section, for our primary analysis, 
we do not distort the galaxy images
by de-projecting, rotating, or scaling them. We
show major-axis aligned stacks in \S\,6 and de-projected,
radially-normalized profiles in an appendix. 
Our results remain qualitatively consistent regardless of this
methodological decision. 
For all analyses, the images were weighted by their \f\, flux so the stack is not 
dominated by a single bright object. 
The \f\ filter covers the full wavelength range of the G141 
grism encompassing the \ha\ emission line. Normalizing
by the \f\ emission hence accounts for very bright \ha\ line emission
without inverse signal-to-noise weighting as normalizing by the
\ha\ emission would.

As a consequence of the grism's low spectral resolution,
we have to account for the blending of emission lines. 
With a FWHM spectral resolution of $\sim100$\,\AA\,, 
 \ha\,$\lambda6563$\AA\, and [N\,{\sc\,ii}]\,$\lambda6548+6583$\AA\, are blended.
To account for the contamination of \ha\, by [N\,{\sc\,ii}],
we scale the measured flux down by a factor of 1.15
(Sanders {et~al.} 2015) and adopt this quantity as the \ha\, flux. 
This is a simplistic correction as [N\,{\sc\,ii}]/\ha\, 
varies between galaxies (e.g. {Savaglio} {et~al.} 2005; {Erb} {et~al.} 2006b; {Maiolino} {et~al.} 2008; {Zahid} {et~al.} 2013; {Leja} {et~al.} 2013; {Wuyts} {et~al.} 2014; Sanders {et~al.} 2015; Shapley {et~al.} 2015) as well as radially
within galaxies (e.g. {Yuan} {et~al.} 2011; {Queyrel} {et~al.} 2012; Swinbank {et~al.} 2012; {Jones} {et~al.} 2013, 2015; F{\"o}rster~Schreiber {et~al.} 2014; {Genzel} {et~al.} 2014b; {Stott} {et~al.} 2014).
{Stott} {et~al.} (2014) find a range of metallicity gradients 
$-0.063<\Delta Z/\Delta r<0.073\,{\rm dex\, kpc}^{-1}$, with the median 
of $\sim0$ (no gradient)
for 20 typical star-forming galaxies at $z\sim1$.
Hence, we choose to adopt a single correction factor
so as not to introduce systematic uncertainties into the data.

Additionally, \ha\,$\lambda6563$\AA\, and
[S\,{\sc\,ii}]\,$\lambda\lambda 6716,6731$\AA\, are resolved 
but are separated by only $\sim3$ resolution elements. 
In this study, we are concerned primarily with the radial 
distribution of \ha\, emission. In order to prevent [S\,{\sc\,ii}]
from adding flux at large radii, we mask the region of the 2D spectrum 
redward of \ha\, where [S\,{\sc\,ii}] emission could contaminate the \ha\, maps.

Galaxies are centered according to the 
light-weighted center of their \f\, flux distribution.
Given that the \f\, can be used as a proxy for stellar mass, 
we chose to center the galaxies according to their \f\, center
as our best approximation of centering them according to stellar mass.
While the \f\, centroid will not always be the exact center of mass,
it is a better estimate than our other option, the \ha\, centroid. 
We measure the centroid of the \f\, images as the flux-weighted 
mean pixel in the x- and y- directions independently with 
an algorithm similar to the iraf task imcntr.
We shift the \f\, image 
with sub-pixel shifts using damped sinc interpolation.
The G141 image is shifted with the same shifts. 
To center the \ha\, map requires only a geometric mapping in the 
spatial direction of 2D grism spectrum. In the spectral direction, 
however, the redshift of a galaxy and the spatial distribution of its 
\ha\, are degenerate. 
As a result, the uncertainty in the spectral direction of the \ha\, 
maps is $\sim0.5$\,pixels (see  {Brammer} {et~al.} 2012a).

\begin{figure*}[hbtf]
\centering 
\includegraphics[width=\textwidth]{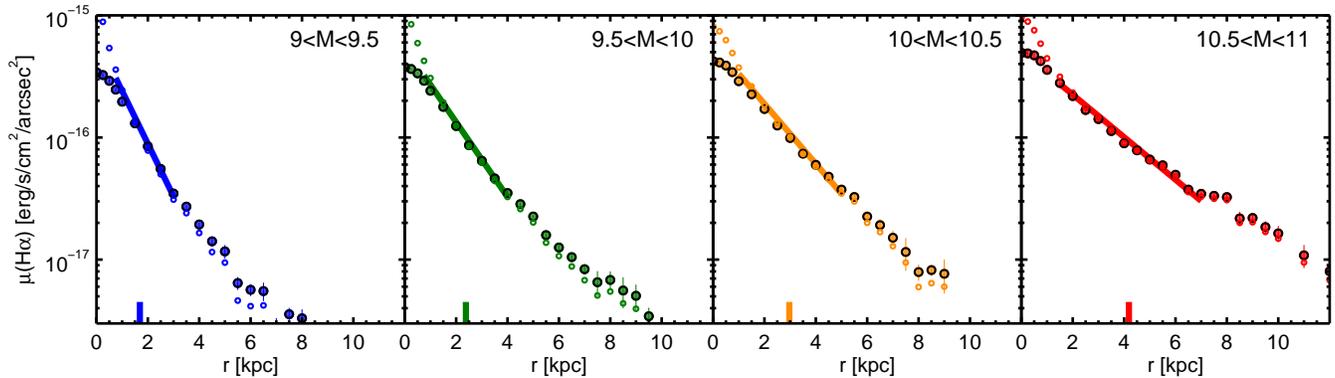}
\caption{The average radial distribution of \ha\, emission in galaxies in 
bins of stellar mass indicated at the top of each panel. 
The filled circles show the radial profiles measured directly
from the stacked \ha\, images. The open circles show the profiles corrected
for the effect of the PSF. 
The lines show the best fit exponentials for $0.5r_s<r<3r_s$ 
to the PSF-corrected profiles. 
There appears to be some excess flux over a pure exponential at 
small and large radii. 
The short vertical lines show the corresponding \ha\ effective radii. 
\label{fig:msprofs}}
\end{figure*}

To simultaneously address these problems, we apply an 
asymmetric double pacman mask to the \ha\, maps.
This mask is shown applied to the stack in Fig.\,\ref{fig:stack}.
The mask serves three purposes. First, it masks the [S\,{\sc\,ii}] emission
which otherwise could masquerade as \ha\, flux at large radii.
Second, it mitigates the effect of the redshift-morphology degeneracy
by removing the parts of the \ha\, distribution that would be most 
affected. Third, it reduces the impact of imperfect 
stellar continuum subtraction by masking the portion of the
spectrum that would be most afflicted. 

A mask was also created for each galaxy's \f\, image to cover pixels
that are potentially affected by neighboring objects. 
This mask was constructed from the 3D-HST photometric data products.
SExtractor was run on the combined $J_{F125W}/H_{F140W}/H_{F160W}$ 
detection image (see  {Skelton} {et~al.} 2014). 
Using the SExtractor segmentation map, we flagged all 
pixels in a postage stamp belonging to other objects and masked them. 
For both \ha\, and \f\, a bad pixel mask is created for known bad or missing 
pixels as determined from the data quality extensions of the 
fits files.

The final mask for each \ha\ image is comprised of the union 
of three separate masks:
1) the bad pixel mask,  
2) the asymmetric double pacman mask, and 
3) the contamination mask (see previous section).
A final \f\, mask is made from the combination of two separate masks
1) the bad pixel mask and 2) the neighbor mask.
The \ha\, and \f\, images are multiplied by these masks before they are summed.
Summing the masks creates what is effectively a weight map 
for the stacks. The raw stacks are divided by this weight map to create
the final exposure-corrected stacked images. 

\subsection{Surface brightness profiles}
The stacked \ha\ image for galaxies with 
$10^{10}<\textrm{M}_*<10^{10.5}$ is shown in Fig.\,\ref{fig:stack}.
With hundreds of galaxies, 
this image is very deep and we can trace the
distribution of \ha\, out to large radii ($\sim10$\,kpc).
To measure the average radial profiles of the \ha\, and \f\, emission, 
we compute the surface brightness as a function 
of radius by measuring the mean flux in circular apertures. 
We checked that the total flux in the stacks matched 
the \ha\, and \f\, fluxes in our catalogs. 
We compute error bars on the radial profiles by bootstrap 
resampling the stacks and in general, we cut off the profiles 
when $S/N<2.5$. 
The \ha\, profile for the example stack is shown in Fig.\,\ref{fig:stack}.
Before moving on to discussing
the trends in the observed radial profiles, we note two additional 
corrections made to them. 

First, we correct the continuum model used to create the \ha\, maps.
This continuum model goes out to the edge of the segmentation map of each
galaxy, which typically encompasses $\gtrsim95$\% of the light. 
We subtract the remaining continuum flux by correcting the continuum model
to have the same spatial distribution as the broad band light. 
The \f\ filter covers the same wavelength range as the 
G141 grism. Therefore, the radial distribution of \f\ emission reflects the true 
radial distribution of continuum emission. We derive a correction factor
to the continuum model of each stack by fitting a second degree polynomial to 
the radial ratio of the \f\ stack to the stacked continuum model.
This continuum correction is $<20\%$ at all radii in the profiles 
shown here.

Second, we correct the radial profiles for the effect of the PSF. 
Compared to typical ground-based observations, 
our space-based PSF is narrow and relatively stable.  
We model the PSF using Tiny Tim ({Krist} 1995) 
and interlacing the model PSFs in the same way as the data.
The FWHM is 0.14", which corresponds to $\sim1$\,kpc at $z\sim1$. 
Although this is small, it has an effect, particularly by blurring
the centers of the radial profiles. 
Images can be corrected using a deconvolution algorithm.
However, there are complications with added noise in low S/N regions and 
no algorithm perfectly reconstructs the intrinsic light distribution
(see e.g. {van Dokkum} {et~al.} 2010). We instead employ 
the algorithmically more straight-forward method of {Szomoru} {et~al.} (2010). 
This method takes advantage of the GALFIT
code which convolves models with the PSF to fit galaxy light
distributions ({Peng} {et~al.} 2002). 
We begin by fitting the stacks with S\'{e}rsic (1968) models using GALFIT
({Peng} {et~al.} 2002).
These  S\'{e}rsic  fits are quite good and the images show small residuals.
We use these fit parameters to create an unconvolved model.
To account for deviations from a perfect  S\'{e}rsic  fit, we add the residuals
to this unconvolved image. Although the residuals are still convolved 
with the PSF, this method has been shown to reconstruct the true 
flux distribution even when the galaxies are poorly fit by a S\'{e}rsic profile
({Szomoru} {et~al.} 2010). It is worth noting again that the residuals in these
fits are small so the residual-correction step in this procedure is not 
critical to the conclusions of this paper.


\section{The distribution of \ha\, as a function of stellar mass and radius}
 \begin{figure}[hbtf]
\centering 
\includegraphics[width=0.5\textwidth]{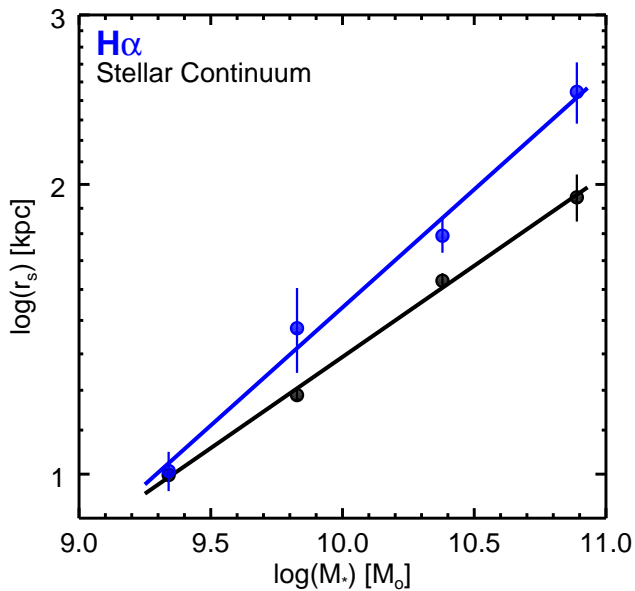}
\caption{Size-mass relations for \ha\, ($r_{H\alpha}-M_*$) 
stellar continuum ($r_*-M_*$). 
The size of star forming disks traced by \ha\, increases with stellar 
mass as $r_{H\alpha}\propto M^{0.23}$. At low masses, 
$r_{H\alpha}\sim r_{*}$, as mass increases the disk scale length of \ha\, 
becomes larger than the stellar continuum emission as
$r_{H\alpha}\propto r_{*}\,M_*^{0.054}$.
Interpreting \ha\, as star formation and stellar continuum as stellar mass, 
this serves as evidence that on average, galaxies are growing larger 
in size due to star formation. 
\label{fig:mass_size}
}
\end{figure}

\begin{figure*}[hbtf]
\centering 
\includegraphics[width=\textwidth]{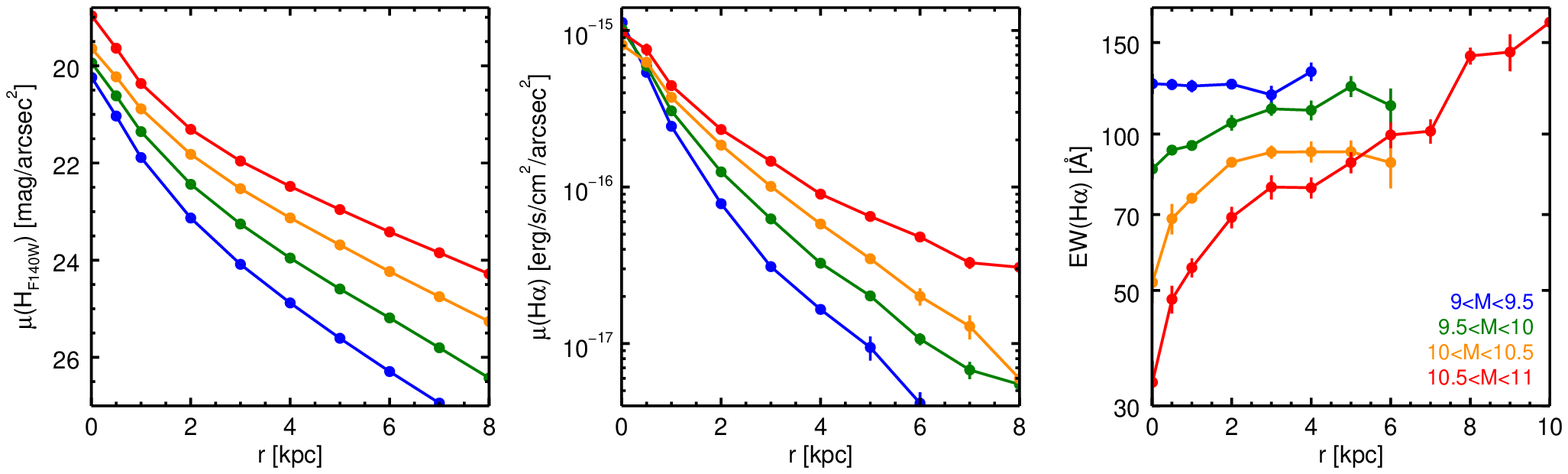}
\caption{Average radial surface brightness profiles 
of \f\ (left), \ha\ (center), and average radial \ha\, equivalent width profile (\ew) (right)
in galaxies as a function of stellar mass.
The radial \ew\ profile is the quotient of the \ha\ 
and stellar continuum profiles, providing a comparison between the spatial distribution 
of \ha\, and stellar continuum emission. At low masses the \ew\ profile is flat. As mass 
increases EW(\ha) rises increasingly steeply from the center, showing, in agreement 
with the larger disk scale lengths of Fig.\,\ref{fig:mass_size}, that the \ha\, has a
more extended distribution than the existing stellar continuum emission. 
\label{fig:mProfs}}
\end{figure*}

The structure of galaxies (e.g.  {Wuyts} {et~al.} 2011a; van~der Wel {et~al.} 2014a)
and their sSFRs (e.g.  {Whitaker} {et~al.} 2014)
change as a function of stellar mass. This means that both 
where a galaxy is growing and how rapidly it is growing depend 
on how much stellar mass it has already assembled. In this 
section, we investigate where galaxies are building stellar mass 
by considering the average radial distribution of \ha\, emission 
in different mass ranges. 

To measure the average spatial distribution of \ha\, during 
this epoch from $z=1.5-0.7$, we create mean \ha\, images by stacking
the \ha\, maps of individual galaxies as described in \S\,3.3. 
The stacking technique employed in this paper serves
to increase the S/N ratio, enabling us to trace the profile of \ha\,
to large radii. An obvious disadvantage is that the \ha\, distribution
is known to be different for different galaxies. As an example,
the \ha\, maps of the galaxies shown in Fig.\,\ref{fig:exmaps} are 
quite diverse, displaying a 
range of sizes, surface densities, and morphologies. 
Additionally, star formation in the early universe often appears to be 
clumpy and stochastic.
Different regions of galaxies light up with new stars for short periods of time. 
These clumps, while visually striking, make up a small fraction 
of the total star formation at any given time. Only $10-15$\% of star formation 
occurs in clumps while the remaining $85-90$\% of star formation occurs in a 
smooth disk or bulge component ({F{\"o}rster Schreiber} {et~al.} 2011b; {Wuyts} {et~al.} 2012, 2013). 
Stacking \ha\, smoothes over the short-timescale stochasticity to reveal the  
time-averaged spatial distribution of star formation. 

Fig.\,\ref{fig:msprofs} shows the radial surface brightness profiles
of \ha\, as a function of stellar mass. 
The first and most obvious feature of these profiles is that
the \ha\, is brightest in the center of these galaxies:
the radial surface brightness of \ha\, rises monotonically 
toward small radii. The average distribution of ionized gas 
is not centrally depressed or even flat, it is centrally peaked. 
This shows that there is substantial on-going star formation 
in the centers of galaxies at all masses at $z\sim1$.

With regard to profile shape, in log(flux)-linear(radius) space, 
these profiles appear to 
be nearly linear indicating they are mostly exponential. 
There is a slight excess at small and large radii compared to an exponential profile.
However, the profile shape is dependent on the
stacking methodology: if the profiles are deprojected and
normalized by their effective radius
(as derived from the \f\ data) they are closer to exponential (see appendix).
We do not use these normalized profiles as the default in the analysis, as it is
difficult to account for the effects of the PSF.

We quantify the size of the ionized gas distribution in two ways:
fitting exponential profiles and S\'{e}rsic models. 
For simplicity, we measure the disk scale lengths ($\equiv r_s$) of the ionized gas 
by fitting the profiles with an exponential between $0.5r_s<r<3r_s$. 
These fits are shown in Fig.\,\ref{fig:msprofs}.
It is clear that over the region $0.5r_s<r<3r_s$ the \ha\, distribution 
is reasonably well-approximated by an exponential. 
Out to $5r_s$, $\sim90\%$ of the \ha\, can be accounted for by this 
single exponential disk fit. 
This implies that most of the \ha\, lies in a disk.

The scale length of the exponential disk fits increases with mass from 1.3\,kpc 
for $9.0<M_*<9.5$ to 2.6\,kpc for $10.5<M_*<11.0$. With $r_e=1.678r_s$, 
this corresponds to effective (half-light) radii of 2.2\,kpc and 4.4\,kpc respectively. 
We fit the size-mass relation of the ionized gas disks 
($r_{H\alpha}-M_*$) with:
 \begin{equation}
r_{H\alpha}(m_*)=1.5m_*^{0.23}
\end{equation}
where $m_*=M_*/10^{10}M_\odot$. Fitting the \f\, surface brightness profiles in the same 
way shows the
exponential disk scale lengths of the stellar continuum emission 
vs. the ionized gas. We parameterize this comparison in terms of the 
stellar continuum size: 
\begin{equation}
r_*(m_*)=1.4m_*^{0.18}
\end{equation}
\begin{equation}
r_{H\alpha}(m_*,r_*)=1.1\,\, r_*\,\,(m_*^{0.054})
\end{equation}
For $10^9M_\odot<M_*<10^{9.5}M_\odot$, the \ha\, emission has the same disk scale length 
as the \f\, emission. This suggests that the \ha\, emission closely follows
the \f\, emission (or possibly the other way around).
At stellar masses $M_*>10^{9.5}$ the scale length of the 
\ha\, emission is larger than the \f. 
As mass increases, the \ha\, grows increasingly more extended 
and does not follow the \f\, emission as closely.
The size-mass relations for \ha\, and \f\, are shown in Fig.\,\ref{fig:mass_size}.

The ionized gas distributions can also 
be parameterized with S\'{e}rsic profiles. 
We fit the observed, PSF-convolved stacks with S\'{e}rsic models 
using GALFIT as described in the previous section. 
The S\'{e}rsic index of each, which reflects the degree of curvature
of the profile, is $1<n<2$ for all mass bins, demonstrating that they
are always disk-dominated. 
The S\'{e}rsic indices and sizes measured with GALFIT are listed
in Table 1. The sizes measured with GALFIT are similar to
those measured using exponential disk fits and exhibit the same
qualitative trends. 

While the bootstrap error bars for each individual method are very small,
$2-4$\%, different methodologies result in systematically different
size measurements.
We derive our default sizes by fitting exponentials to the $0.5r_s<r<3r_s$
region of PSF-corrected profiles. 
Fit the same way, sizes are $10-20$\% larger
when profiles are not corrected for the PSF.
Adopting slightly different fitting regions can also change the sizes by 
$10-20$\%. The GALFIT sizes are $3-15$\% larger. 
With all methods the trends described remain 
qualitatively the same. That is, the effective radius of the \ha\ 
emission is always greater or equal to the effective radius of the \f\ 
and both increase with stellar mass.

\begin{table}
\begin{center}
\caption{Structural Parameters} \label{tab:structPar}
\begin{tabular}{ l r  r  r  r  r r }
\toprule
 & \multicolumn{3}{c}{\ha} & \multicolumn{3}{c}{\f} \\
\cmidrule(r){2-4} \cmidrule(r){5-7}
log(M$_*$) & r$_s$ & r$_e$ &  n  & r$_s$ & r$_e$ &  n \\[2mm] 
\hline
    $9.0<\textrm{log(M}_*)<9.5$ &   1.0 &   1.8 &   1.9 &   1.0 &   1.8 &   1.9 \\
   $9.5<\textrm{log(M}_*)<10.0$ &   1.5 &   2.7 &   1.8 &   1.3 &   2.4 &   1.9 \\
  $10.0<\textrm{log(M}_*)<10.5$ &   1.8 &   3.2 &   1.5 &   1.6 &   3.1 &   1.7 \\
  $10.5<\textrm{log(M}_*)< 11.0$ &  2.6 &   5.1 &   1.7 &   2.0 &   3.9 &   2.1 \\
\hline
\end{tabular}
\\[3mm] 
\end{center}
\textit{Note.} Disk scale length and effective radius in kpc and S\'{e}rsic index for 
\ha\, and \f\, as a function of stellar mass. For an exponential disk (n=1), 
$r_e=1.678r_s$.
\end{table}

The comparison between the radial distribution of \ha\ and \f\ can be seen 
explicitly in their quotient, the radial \ha\ equivalent width (EW(\ha)) profile 
(Fig. \ref{fig:mProfs}), indicating where the \ha\ emission is elevated and 
depressed relative to the \f\ emission.  
The first and most obvious feature is that the normalization of equivalent 
width profiles decreases with increasing stellar mass, 
consistent with spatially-integrated results (Fumagalli {et~al.} 2012) 
and the fact that sSFR declines with stellar mass (e.g.  {Whitaker} {et~al.} 2014). 
Additionally, below a stellar mass of $\textrm{log(M)}_*<9.5$,
the equivalent width profile is flat, at least on the scales of $\sim1$\.kpc resolved
by our data. 
These galaxies are growing rapidly across their disks.
In addition to the overall normalization of the EW decreasing, 
as stellar mass increases the shape of the EW profile changes,
its slope growing steeper. For $9.5<{\rm log(M_*)}<10.0$, \ew\
rises by a factor of $\sim1.3$ from the center to 2r$_e$,
for $10.5<{\rm log(M_*)}<11.0$, it rises by $\gtrsim3$.
At low masses, the entire disk is illuminated with new stars; 
at higher masses, the \ha\, is somewhat centrally depressed relative
to the stellar continuum emission. Consistent with the measured size trends,
the radial EW(\ha) profiles show that \ha\, has a similar distribution as the stellar 
continuum emission for $9.0<{\rm log(M_*)}<9.5$; as mass increases \ha\, becomes more
extended and less centrally concentrated than the stellar continuum emission. 

Interpreting \ha\ as star formation and \f\ as stellar mass implies 
that star formation during the epoch $0.7<z<1.5$ is building galaxies 
from the inside-out as discussed in \S\,7.3.

\section{The radial distribution of \ha\, across the star forming sequence}

\begin{figure}[hbtf]
\centering 
\includegraphics[width=0.45\textwidth]{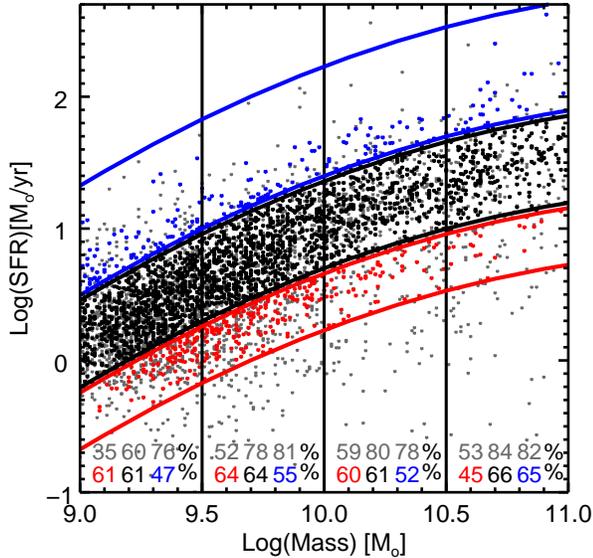}
\caption{We investigate the spatial distribution of star formation in galaxies 
across the SFR(UV+IR)-\m\ plane. To do this, we stack the \ha\ maps 
of galaxies on the star forming sequence main sequence (black) and compare 
to the spatial distribution of \ha\ in galaxies above (blue) and below 
(red) the main sequence. The parent sample is shown in gray. 
The fractions of the total parent sample above the \ha\ flux and 
extraction magnitude limit are listed at the bottom in gray. 
As expected, we are significantly less
complete at low masses, below the main sequence. 
About one third of selected galaxies 
are thrown out of the stacks due to contamination of their spectra by other sources in 
the field. Of the galaxies above the flux and extraction limits, the
fractions remaining as part of the the final selection are listed and shown in
blue/black/red and respectively. 
\label{fig:bins}}
\end{figure}

\begin{figure*}[hbtf]
\centering
\includegraphics[width=\textwidth]{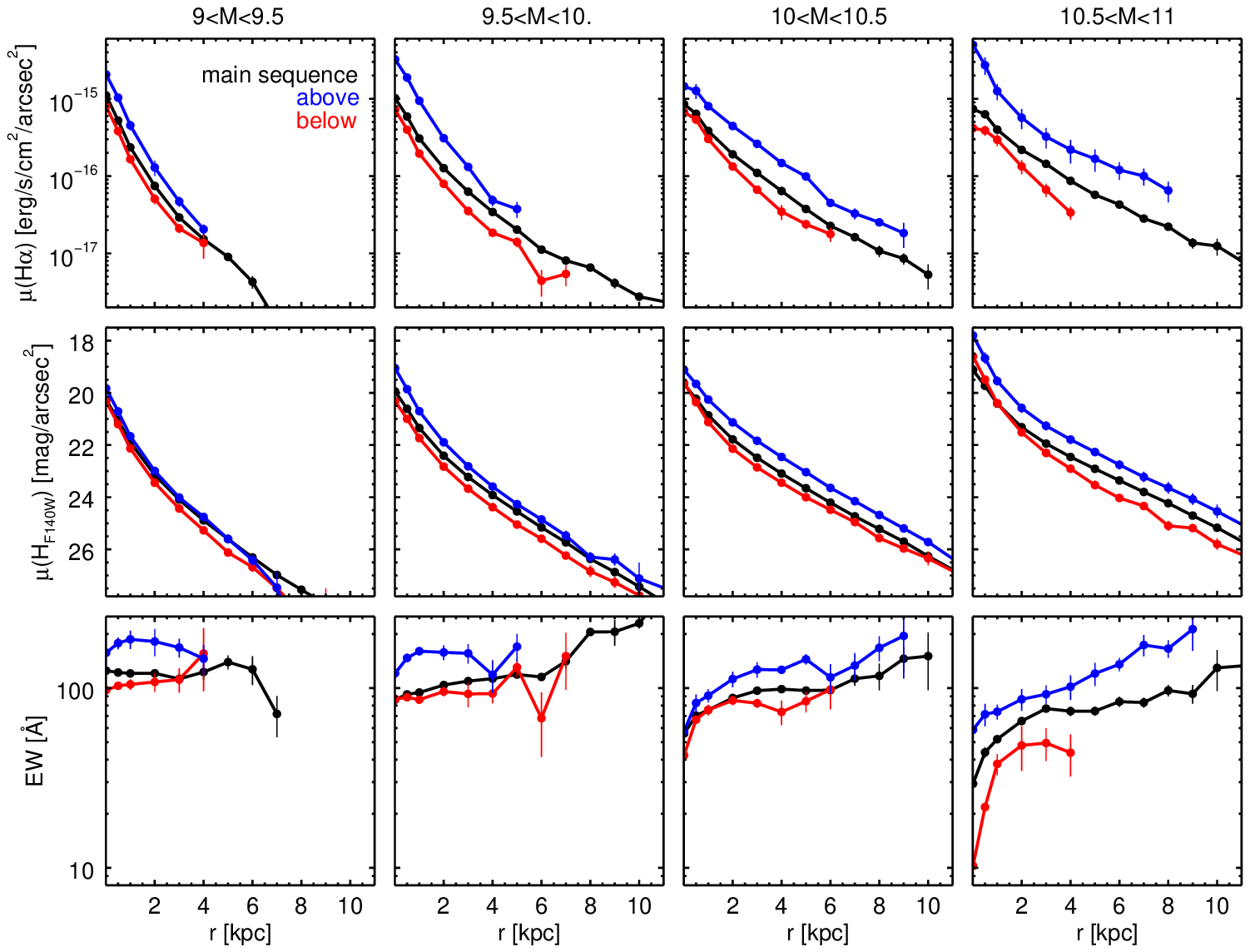}
\caption{Radial surface brightness profiles of \ha, \f, and their ratio 
EW(\ha) as a function of \m\, and SFR. The colors delineate position 
with respect to the star forming `main sequence':
above (blue), on (black), and below (red). 
Above the star forming main sequence, the \ha\, (as well as the \f\ and EW(\ha)
is elevated at all radii. 
Below the star forming main sequence, the \ha\, is depressed at all radii.
The average radial profiles are always centrally peaked in \ha\, and never 
centrally peaked in EW(\ha).
\label{fig:obsprofs}}
\end{figure*}

In the previous section, we showed how the radial distribution of star 
formation depends on the stellar mass of a galaxy. Here we show how it 
depends on the total star formation rate at fixed mass. In other words, 
we show how it depends on a galaxy's position in the SFR-\m\ plane 
with respect to the star forming main sequence. 
(The star forming 'main sequence' is an observed locus of points
in the SFR-\m\ plane  {Brinchmann} {et~al.} 2004; {Zheng} {et~al.} 2007; {Noeske} {et~al.} 2007; {Elbaz} {et~al.} 2007; {Daddi} {et~al.} 2007; {Salim} {et~al.} 2007; {Damen} {et~al.} 2009; {Magdis} {et~al.} 2010; {Gonz{\'a}lez} {et~al.} 2010; Karim {et~al.} 2011; {Huang} {et~al.} 2012; {Whitaker} {et~al.} 2012, 2014)

\subsection{Definition of the Star Forming Main Sequence}

We define the star forming sequence according to the results of
{Whitaker} {et~al.} (2014),
interpolated to $z=1$. 
The slope of the relation between SFR and \m\, decreases
with \m, as predicted from galaxy growth rates derived from
the evolution of the stellar mass function ({Leja} {et~al.} 2015), reflecting 
the decreased efficiency of stellar mass growth at low and high masses. 
{Whitaker} {et~al.} (2014) find that the observed scatter is a constant $\sigma=0.34$\,dex 
with both redshift and \m. 

We investigate where `normal' star-forming galaxies were forming their 
stars at this epoch by determining the radial distribution of \ha\, in galaxies
on the main sequence. We elucidate how star formation is enhanced and 
suppressed in galaxies by determining where star formation is "added"
in galaxies above the main sequence and "subtracted" in galaxies below
the main sequence. 
To determine where star formation is occurring in 
galaxies in these different 
regions of the SFR-\m\, plane, we stack \ha\, maps as a 
function of mass and SFR. We define the main sequence as 
galaxies with SFRs $\pm1.2\sigma=\pm0.4$\,dex from the 
{Whitaker} {et~al.} (2014) main sequence line at $z\sim1$.
Specifically, we consider galaxies `below', `on', or `above'
the star forming main sequence to be the regions
[-0.8,-0.4]\,dex, [-0.4,+0.4]\,dex, or [+0.4,+1.2]\,dex with respect
to the main sequence line in the SFR-\m\ plane. To define these regions
consistently we normalize the SFRs of all galaxies to $z\sim1$ using
the redshift evolution of the normalization of the star forming 
sequence from {Whitaker} {et~al.} (2012). 
These definitions are shown pictorially by Fig. \ref{fig:bins}
in red, black, and blue respectively. 
We imposed the +1.2\,dex upper limit above the main sequence 
so the stacks wouldn't be dominated by a single, very bright galaxy. 
We impose the -0.8\,dex due to the \ha\, flux-driven completeness limit. 
Fig. \ref{fig:bins} also shows which galaxies were actually used in the stacks. 
Our broad band magnitude extraction limit and \ha\, flux limit manifest themselves 
as incompleteness primarily at low masses and SFRs as reflected in the gray 
numbers and filled symbols. 

We adopted this $\pm1.2\sigma$ definition of the main sequence 
to enable us to probe the top and bottom 10\% of star formers 
and ferret out differences between galaxies growing very rapidly, 
very slowly, and those growing relatively normally. 
According to our definition ($\pm1.2\sigma$), the `Main Sequence' accounts 
for the vast majority of galaxy growth. It encompasses 80\% of 
UVJ star-forming galaxies and 76\% of star formation. 
The star forming main sequence
is defined by the running median star formation rate of galaxies as a 
function of mass. The definition is nearly identical when the mode is used
instead, indicating that it defines the most common rate of 
growth. 
While we left 20\% of star-forming galaxies to probe the extremes of 
rapid and slow growth, only 7\% of these galaxies live above the main sequence
and nearly double that, 13\%, live below it. 
This is a manifestation of the fact that the distribution of star formation rates at a 
given mass is skewed toward low star formation rates. 
Counting galaxies, however, 
understates the importance of galaxies above the main sequence to 
galaxy evolution because they are building stellar mass so rapidly. 
Considering instead the contribution to the total star formation
budget at this epoch, galaxies above the main sequence account
for $>20\%$ of star formation
while galaxies below the main sequenceonly account for $<3\%$.

\subsection{Results}

\begin{figure}[hbtf]
\centering
\includegraphics[width=0.5\textwidth]{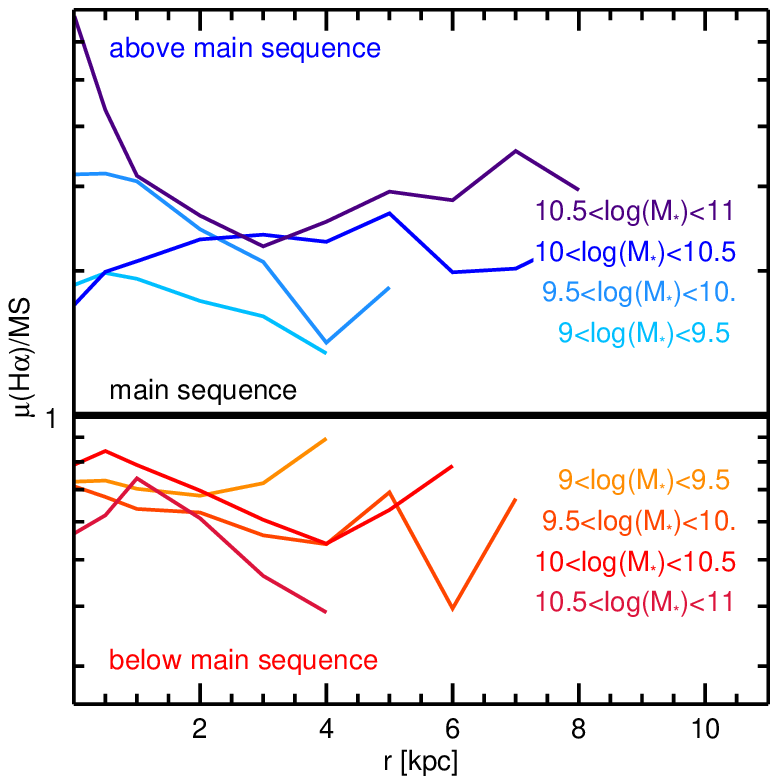}
\caption{Radial profiles of \ha\, as a function of mass normalized by 
the main sequence radial profile (MS). 
Above the star forming main sequence, 
the \ha\, is elevated at all radii (blue hues). 
Below the star forming main sequence,
the \ha\, is depressed at all radii (red hues).
\label{fig:divsum}}
\end{figure}

One of the primary results of this paper is shown in 
Fig.\,\ref{fig:obsprofs}: the radial distribution
of \ha\, on, above and below the star forming main sequence. 
Above the main sequence, \ha\, is elevated at all radii. 
Below the main sequence, \ha\, is depressed at all radii. 
The profiles are remarkably similar above, on, and below the 
main sequence -- a phenomenon that can be referred to
as `coherent star formation', in the sense that the
offsets in the star formation rate are spatially-coherent.
As shown in and Fig.\,\ref{fig:divsum},
the offset is roughly a factor of 2 and nearly 
independent of radius: at $r<2$\,kpc the mean offset is a factor of 2.2, 
at $3<r<5$\,kpc it is a factor of 2.1. 
Above the main sequence at the highest masses where we have
the signal-to-noise to trace the \ha\, to large radii, we 
can see that the \ha\, remains enhanced by a factor of $\gtrsim2$
even beyond 10\,kpc. 
The most robust conclusion we can draw from the radial profiles of \ha\, 
is that star formation from $\sim2-6$\,kpc is enhanced in galaxies 
above the main sequence and suppressed in galaxies below the main 
sequence (but see \S\,7.4 for further discussion). 

We emphasize that the SFRs used in this paper were derived from UV+IR emission,
These star formation rate indicators are measured independently from the \ha\, flux. 
Thus, it is not a priori clear that the \ha\ emission is enhanced or depressed
for galaxies above or below the star forming main sequence as
derived from the UV+IR emission. 
The fact that it is implies that the scatter in the star forming 
sequence is real and caused by variations in the star formation
rate (see \S\,7.4).

In the middle panels of Fig.\,\ref{fig:obsprofs} we show the radial profiles
of \f\, emission as a function of \m\, above, on, and below the star 
forming main sequence. As expected, we find
that the average sizes and S\'{e}rsic 
indices of galaxies increase with increasing stellar mass.
Disk scale lengths of \ha\ and \f\ are listed in Table 2. 
At high masses, we find that above and below the main sequence, the \f\ is  
somewhat more centrally concentrated than on the main sequence 
(consistent with {Wuyts} {et~al.} 2011a; {Lang} {et~al.} 2014, Whitaker et al. in prep), possibly
indicating more dominant bulges below and above the main sequence.
We note that these trends are less obvious at lower masses.
Furthermore, as one would expect, 
the mass to light ratio decreases with sSFR because young stars
are brighter than old stars. Therefore, at fixed mass,
galaxies above the main sequence have brighter \f\ stellar continuum
emission and galaxies below the main sequence have fainter \f\ emission.

\begin{table*}
\begin{center}
\caption{Disk scale lengths of \ha\ and stellar continuum emission below, on, and above the star forming main sequence} \label{tab:sizes}
\begin{tabular}{ l r  r  r  r  r r }
\toprule
 & \multicolumn{3}{c}{r$_s$(\ha) [kpc]} & \multicolumn{3}{c}{r$_s$(\f) [kpc]} \\
\cmidrule(r){2-4} \cmidrule(r){5-7}
log(M$_*$) & below & MS & above  & below & MS & above \\[2mm] 
\hline
   $9.0<\textrm{log(M}_*)<9.5$  & $ 1.43 \pm 0.28$  & $ 1.24 \pm 0.06$  & $ 1.12 \pm 0.06$  & $ 1.17 \pm 0.03$  & $ 1.24 \pm 0.01$  & $ 1.17 \pm 0.03$ \\ 
  $9.5<\textrm{log(M}_*)<10.0$  & $ 1.44 \pm 0.07$  & $ 1.68 \pm 0.02$  & $ 1.20 \pm 0.15$  & $ 1.46 \pm 0.03$  & $ 1.51 \pm 0.01$  & $ 1.27 \pm 0.09$ \\ 
 $10.0<\textrm{log(M}_*)<10.5$  & $ 1.90 \pm 0.14$  & $ 1.99 \pm 0.05$  & $ 1.95 \pm 0.08$  & $ 1.78 \pm 0.08$  & $ 1.83 \pm 0.02$  & $ 1.82 \pm 0.09$ \\ 
$10.5<\textrm{log(M}_*)< 11.0$  & $ 1.68 \pm 0.11$  & $ 2.60 \pm 0.08$  & $ 3.14 \pm 0.49$  & $ 1.57 \pm 0.02$  & $ 2.22 \pm 0.05$  & $ 1.86 \pm 0.13$ \\ 
\hline
\end{tabular}
\\[3mm] 
\end{center}
* For an exponential disk (n=1), the half-light radius is $r_e=1.678r_s$.
\end{table*}

In the bottom panels of Fig.\,\ref{fig:obsprofs} we show the radial \ew\ profiles. 
The most obvious 
feature of these profiles is that EW(\ha) is \textit{never} centrally peaked. 
\ew\ is always flat or centrally depressed, indicating 
the \ha\ is always equally or less centrally concentrated than 
the the \f\ emission. 
Above the main sequence, the \ew\ is elevated at all radii. 
Below the main sequence, the \ew\ is depressed at most radii.
These trends are discussed more extensively in \S\,7.4-5, where
we convert the \ew\ profiles to sSFR profiles.



\section{Effects of orientation}
\begin{figure*}[hbtf]
\centering
\includegraphics[width=0.9\textwidth]{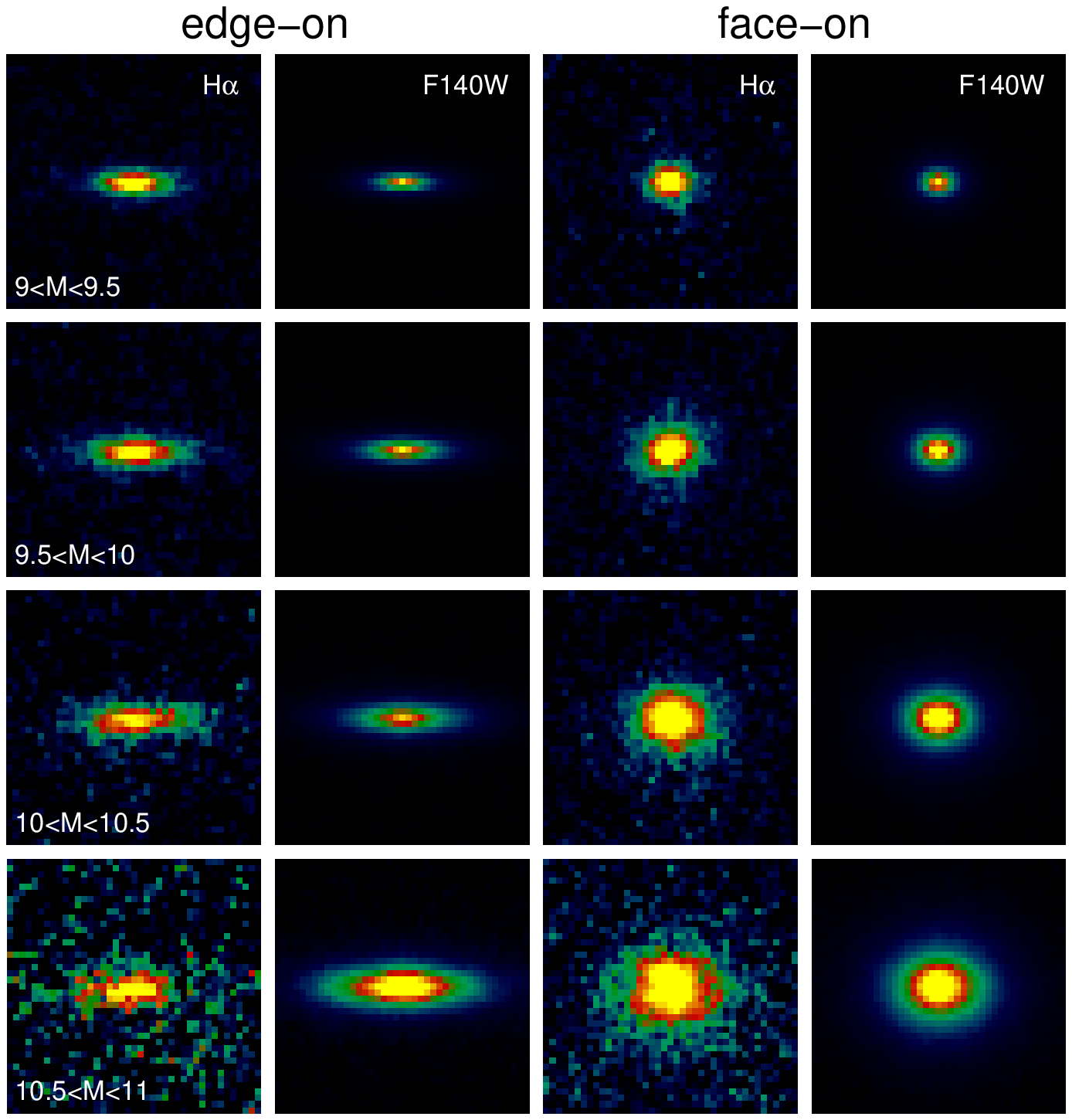}
\caption{Stacks of galaxies after rotating them so their
major axes are aligned, for the
20\,\% of galaxies with the lowest ellipticities (``face-on'') and the 20\,\% of
galaxies with the highest ellipticiticies (``edge-on''). 
The lowest ellipticity stacks are nearly round and the highest ellipticity stacks are highly
flattened with $a/b \approx 0.3$,
consistent with viewing disks under different projections.  The \ha\ stacks
are remarkably similar to the \f\ stacks, demonstrating that the \ha\ emission
is aligned with the \f\ emission at all masses.
\label{fig:rotStacks}}
\end{figure*}

In the previous sections we analyzed average images and radial profiles of 
\ha\, emission with galaxies stacked as they were oriented on the detector. 
This methodology has the advantage that it
allows for better control of systematics. In particular,
we can effectively subtract the  continuum out to large radii as
we can use the radial distribution of the \f\, flux to correct for the $\leq5$\% of flux missing
from the continuum models.  
A galaxy's position angle on the detector, however, is arbitrary and has no 
physical meaning. 

Here we present stacks of galaxies rotated to be aligned along the major axis, as measured
from the continuum emission.  This is an important test of the idea that
the \ha\ emission originates in disks that are aligned with the stellar
distribution: in that case
these rotated \ha\ stacks should have similar axis ratios as the rotated \f\ stacks.
We divide the galaxies into the same mass bins 
as in the previous sections, and compare the most face-on vs. the most edge-on
galaxies. The position angle and projected axis ratio ($q=B/A$) of each galaxy 
is measured from
its \f\, image using GALFIT ({Peng} {et~al.} 2002). 
We rotate the \f\, and \ha\, images according to their \f\, position angle 
to align them along the major axis. 
In each mass bin, we then create face- and edge-on stacks from the galaxies with
the highest and lowest 20\% in projected axis ratio, respectively. 

The distribution of projected axis ratios is expected to be broad if 
most galaxies are disk-dominated (see, e.g., van~der Wel {et~al.} 2014b). 
If we interpret the galaxy images as disks under different orientations, 
we would expect the stacks of galaxies with the highest 20\% of projected 
axis ratios to have an average axis ratio of
$\sim 0.9$ and the stacks of galaxies with the lowest 20\% of 
projected axis ratios to be flattened with average axis ratios of $\sim 0.3$
(see van~der Wel {et~al.} 2014b).
As shown in Fig.\ \ref{fig:rotStacks}
the rotated \f\ stacks are
consistent with this expectation. Furthermore, the rotated
\ha\ stacks are qualitatively very similar to the rotated \f\ stacks,
which means that the \ha\ emission is aligned with that of the stars.

For the edge-on stacks, we measure the flattening of the \ha\, 
emission and compare it to that of the \f\, emission.  In the four mass
bins, from low mass to high mass, we find 
$q(H\alpha)=[0.29\pm0.02,0.32\pm0.03,0.31\pm0.02,0.37\pm0.02]$ and
$q(H_{F140W})=[0.28\pm0.01,0.27\pm0.01,0.29\pm0.01, 0.34\pm0.01]$ respectively,
where the errors are determined from bootstrap resampling.
We find that the average axis ratio of \f\ emission is $q(H_{F140W})=0.295 \pm 0.005$ and
$q(H\alpha)=0.323 \pm 0.011$.  We conclude that 
the \ha\, is slightly less flattened than the \f\ emission, but the difference is
only marginally significant.

There are physical reasons why \ha\ can have an intrinsically larger 
scale height than the \f\ emission.
Given that outflows are ubiquitous in the $z\sim2$ universe 
(e.g. {Shapley} {et~al.} 2003; {Shapiro} {et~al.} 2009; {Genzel} {et~al.} 2011; {Newman} {et~al.} 2012; {Kornei} {et~al.} 2012; F{\"o}rster~Schreiber {et~al.} 2014; {Genzel} {et~al.} 2014a),
it is possible that the \ha\ would have a larger scale height due winds driving
ionized gas out of the plane of the stellar disk.  Furthermore,
attenuation towards HII regions could be more 
severe in the midplane of the disk than outside of it. This would result in \ha\ 
emission being less concentrated around the plane of the disk, giving a larger 
scale height. Finally, the gas disks and the stellar disks can be
misaligned. The fact that the edge-on \ha\ and \f\ stacks are so similar
shows that all these effects are small.

At a more basic level, an important implication of the similarity of the \ha\ stacks and
the \f\ stacks is that it directly shows that we are not stacking noise peaks.
If we were just stacking noise, a stack 
of galaxies flattened in \f\ would not be flattened in \ha\ because the noise would
not know about the shape of the \f\ emission.  It is remarkable that this holds
even for the lowest mass stack,
which contains the galaxies
with the lowest \ha\ S/N ratio as well as the smallest
disk scale lengths.


\section{Discussion}
Thus far, we have only discussed direct observables:
\f\ and \ha. In this Section 
we explicitly interpret the radial profiles of \ha\ 
as radial profiles of star formation and the
radial profiles of \f\ as radial profiles of stellar surface density.

\subsection{Interpreting \ha\ and \f\ as SFR and Mass}
In \S\,4 and \S\,5, we showed the radial distribution of \ha, \f, and \ew.
\ha\ emission is typically used as a tracer of star formation, \f\ 
(rest-frame optical) emission as a proxy for stellar mass, 
and \ew\, for the specific star formation rate (sSFR)
(e.g. {F{\"o}rster Schreiber} {et~al.} 2011a; {Wuyts} {et~al.} 2013; {Genzel} {et~al.} 2014b; {Tacchella} {et~al.} 2015b, 2015a). 
We do the same here to gain more physical insight into the observed profiles.
If we assume that \ha\ traces star formation and \f\ traces stellar mass,
the profiles can be scaled to these physical quantities using the integrated 
values.
To derive mass surface density profiles, we ignore M/L gradients and apply  
the integrated \m$/L_{F140W}$ as a constant scale factor at all radii. 
Similarly, to derive star formation surface density profile, 
we ignore radial dust gradients and scale the \ha\, profiles 
based on the integrated $SFR(UV+IR)/L_{H\alpha}$ ratio.
The sSFR profile is then the quotient of the SFR and \m\ profiles. 
However, there are a number of caveats associated with interpreting 
the \f, \ha, and \ew\ profiles in this manner.

We first assess the assumption that there are no radial gradients
in the SFR/\ha\ ratio.
This assumption can be undermined in four ways: dust,  
AGN, winds, and metallicity, which have opposing effects.
Dust will increase the SFR/\ha\ ratio by obscuring the ionizing 
photons from star forming regions. 
AGN, winds, and higher metallicity will reduce the SFR/\ha\ ratio, as they 
add ionizing photons that do not trace star formation. 
These aspects, and hence the extent to which a scaling from \ha\, to SFR is a good 
assumption, themselves depend on stellar mass and star formation rate.
Dust attenuation is correlated with stellar mass 
(e.g. {Reddy} {et~al.} 2006, 2010; {Pannella} {et~al.} 2009; {Wuyts} {et~al.} 2011b; {Whitaker} {et~al.} 2012; {Momcheva} {et~al.} 2013).
At fixed mass, dust attenuation
is also correlated with star formation rate 
({Wang} \& {Heckman} 1996; {Adelberger} \& {Steidel} 2000; {Hopkins} {et~al.} 2001; {Reddy} {et~al.} 2006, 2010; {Wuyts} {et~al.} 2011b; {Sobral} {et~al.} 2012; {Dom{\'{\i}}nguez} {et~al.} 2013; {Reddy} {et~al.} 2015). 
Within galaxies, dust attenuation is anti-correlated with radius 
(e.g., {Wuyts} {et~al.} 2012), as it depends on the column density. 
This means that SFR and H$\alpha$ should trace each other reasonably well for
low mass galaxies with low star formation rates, and particularly poorly in the
the centers of massive, rapidly star-forming galaxies
({Nelson} {et~al.} 2014; {van Dokkum} {et~al.} 2015).
The same qualitative scalings with mass and star formation
likely apply to the likelihood of 
an AGN being present, outflows, and the contamination of \ha\ by [N\,{\sc\,ii}]. 
That is, AGN are most likely to haunt 
the centers of massive, rapidly star-forming galaxies 
(e.g., {Rosario} {et~al.} 2013; F{\"o}rster~Schreiber {et~al.} 2014; {Genzel} {et~al.} 2014a). 
[N\,{\sc\,ii}]/\ha\, is most likely to be enhanced above the assumed 
value in the centers of massive galaxies (as described in \S3.3).
Shocks from winds may contribute to the \ha\ emission in the central
regions, particularly at high masses 
({Newman} {et~al.} 2012; F{\"o}rster~Schreiber {et~al.} 2014; {Genzel} {et~al.} 2014a).
The takeaway here is that we are relatively confident interpreting 
\ha\, as star formation at low masses, low SFRs, and all profiles 
outside of the center. We are less confident for the centers of the 
radial profiles of massive or highly star-forming galaxies. 

Next, we assess the assumption that there is no radial
gradient in the M/L ratio.
Dust and AGN affect the M/L 
in the same way as SFR/\ha\, although less strongly
(e.g.  {Calzetti} {et~al.} 2000; {Wuyts} {et~al.} 2013; {Marsan} {et~al.} 2015; {Reddy} {et~al.} 2015). 
Galaxies growing inside-out will also have gradients in their
stellar population ages. Since older stellar populations have 
higher M/L ratios, these age gradients translate into M/L gradients. 
Age and dust increase \m/\f\, and AGN decrease it. 
Hence using \f\, as a proxy for \m\, is a fairly 
safe assumption at lower masses where age and dust gradients 
are small and AGN are rare. It is somewhat less certain 
at high masses. 
We also note that the contribution of the 
\ha\, emission to the total \f\, flux is small, $\sim5\%$.

As the \ew\ profile is the quotient of the \ha\ and \f\,
interpreting it as a profile of sSFR is accompanied by the amalgam of
all of the above uncertainties: dust, age, AGN, and metallicity.  
This does not necessarily mean that the sSFR profile
is more uncertain than the
profiles of star formation and mass, as some effects cancel.
In a two component dust model (e.g. {Calzetti}, {Kinney}, \&  {Storchi-Bergmann} 1994; {Charlot} \& {Fall} 2000), 
the light from both stars and HII regions is attenuated by 
diffuse dust in the ISM. The light from the HII regions is attenuated additionally
by dust in the undissipated birth clouds. 
Because the continuum and line emission will be affected equally by 
the diffuse dust, the \ew\ profile will only be affected by the extra 
attenuation toward the stellar birth clouds, not the totality of the dust column.
As a consequence, the effect of dust on the \ew\ profiles is mitigated
relative to the \ha\ profiles. The quantity of extra
attenuation towards HII regions remains a matter of debate
with estimates ranging from none ({Erb} {et~al.} 2006a; {Reddy} {et~al.} 2010) 
to a factor of 2.3 ({Calzetti} {et~al.} 2000; {Yoshikawa} {et~al.} 2010; {Wuyts} {et~al.} 2013) 
and many in between 
(e.g., {F{\"o}rster Schreiber} {et~al.} 2009; {Wuyts} {et~al.} 2011b; Mancini {et~al.} 2011; {Kashino} {et~al.} 2013).
As with the total attenuation, the quantity of extra attenuation toward HII regions
appears to increase with \m\ and SFR ({Price} {et~al.} 2014; {Reddy} {et~al.} 2015).
Reddy et al.\ (2015) find that extra attenuation becomes significant 
at SFR$\sim20{\rm M}_\odot/{\rm yr}$. If true, extra extinction should 
be taken into account for galaxies on the main sequence at the highest masses,
and above the main sequence at ${\rm log(M}_*)>9.5$. 
The issue should be less acute for galaxies with low masses and SFRs. The only
way to definitively resolve this question is to obtain
spatially-resolved dust maps in the future.

\subsection{Star formation in disks}
\begin{figure*}[hbtf]
\centering 
\includegraphics[width=\textwidth]{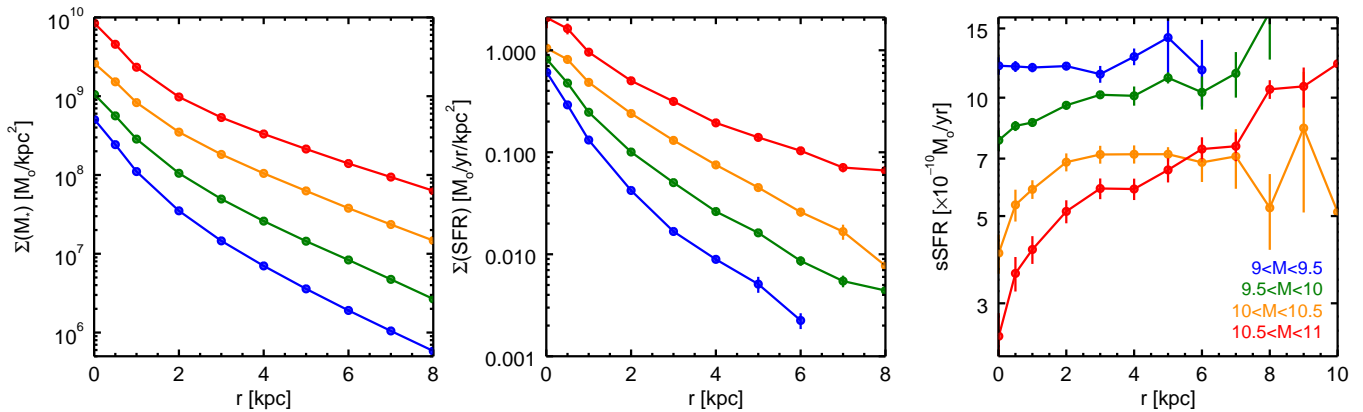}
\caption{Stellar mass surface density(left), star formation surface density(center), and
specific star formation rate (right) as a function of radius and stellar mass. 
These profiles were made by scaling the profiles of \ha, \f, and 
EW(\ha) in Fig.\,\ref{fig:mProfs} to SFR, \m, and sSFR.
Within galaxies, the specific star formation rate rises radially;
the star formation is more extended than the existing stellar mass. 
This is a direct demonstration that galaxies at this epoch grow inside-out.
\label{fig:mphysProfs}}
\end{figure*}

The center panel of Fig.\,\ref{fig:mphysProfs} shows the radial 
distribution of SFR as a function of stellar mass derived by 
scaling the \ha\ profiles to the total SFR(UV+IR). 
The radial distribution of SFR is consistent with 
being disk-dominated: as discussed in \S\,4,
an exponential provides a reasonably good
fit to the profiles and the  S\'{e}rsic  indices are $1<n<2$.
Out to $7r_s$, $\sim85\%$ of the \ha\, can be accounted for by a
single exponential disk fit.  Approximately
$15$\% of the \ha\ emission is in excess above an 
exponential: 5\% from the center ($<0.5r_s$) and 10\% from large radii 
($>3r_s$).

Taken at face value the shape of the stacked \ha\ profiles suggests 
that the star formation during the epoch $0.7<z<1.5$ mostly happens in disks 
with the remainder building central bulges and stellar halos.
In reality, of course, the universe is likely much more complicated. 
Radial dust gradients will make the star formation appear less centrally 
concentrated. 
Stacking galaxies of different sizes will make the star formation appear
more centrally concentrated, as shown in the appendix. 
Additionally, the gas that \ha\, traces can be ionized by physical processes
other than star formation
such as AGN, winds, or shock heating from the halo. 
So with the \ha\ we observe we may also be 
witnessing the growth of black holes, excited gas being driven out of galaxies,
or the shock heating of the inflowing gas that fuels star formation. 

Interestingly, a common
feature of the \ha\, profiles is that they all peak at the center. 
If we interpret the \ha\, as star formation, this means that at all masses, 
galaxies are building their centers. 
Although we caution that shocks from winds and AGN could add
\ha\  (F{\"o}rster~Schreiber {et~al.} 2014; {Genzel} {et~al.} 2014a)
and dust attenuation could subtract \ha\ 
from the centers of the profiles.
That we observe \ha\ to be centrally peaked was not necessarily expected: recently
it was found that some massive galaxies at $z\sim 2$ have H$\alpha$ rings
(see e.g. {Genzel} {et~al.} 2014b; {Tacchella} {et~al.} 2015a), which have been interpreted as
evidence for inside-out quenching.  
We note that our averaged profiles do not exclude
the possibility that some individual galaxies have rings at $z\sim 1$, which are
offset by galaxies with excess emission in the center.

\subsection{Inside-out Growth}
The star formation surface density (as traced by \ha)
is always centrally peaked but the sSFR (as traced by \ew) is
never centrally peaked.
Confirming Nelson {et~al.} (2013) we find that, in general, 
\ew\ is lower in the center than at larger radii.
Confirming {Nelson} {et~al.} (2012), we find that the effective radius of the
\ha\ emission is generally larger than the effective radius of the \f\
emission. This means that the \ha\ emission is more extended
and/or less centrally concentrated than the \f\ emission. 
If \ha\ traces star formation and \f\ traces stellar mass,
these results indicate that galaxies have radial gradients in their specific 
star formation rates: the sSFR increases with radius. 
If the centers are growing more slowly than the outskirts, 
galaxies will build outward, adding proportionally 
more stars at larger radii. 
This suggests that star formation is increasing the size of galaxies.
However, galaxies are still building significantly at their centers
(probably even more than we see due to the effects of dust)
consistent with the fact that 
size growth due to star formation appears to be 
fairly weak (van Dokkum {et~al.} 2013; van~der Wel {et~al.} 2014a; {van Dokkum} {et~al.} 2015).

Additionally, there appears to be a trend in 
$r_s(H\alpha)/r_s(H_{F140W})$ with mass.
Below $3\times10^9\textrm{M}_{\odot}$, the \ha\ and the 
\f\, roughly trace each other: the radial EW profile is flat and 
$r_s(H\alpha)\sim r_s(H_{F140W})$.  
As mass increases, \ha\, becomes 
more extended than the \f\, emission: 
the EW(\ha) profile is increasingly centrally depressed and 
$r_s(H\alpha) > r_s(H_{F140W})$. 
This reflects the natural expectations
of inside out growth and the shape of the sSFR-\m\, relation from both
a physical and an observational standpoint. 

Observationally, our tracers \ha\, and \f\, may trace
somewhat different things as a function of increasing stellar mass. 
At the low mass end, because low mass galaxies have such high sSFRs,
it's possible that the \f\, emission is dominated by light from young 
stars and is not actually a good tracer of stellar mass. This means 
that there may in fact be a difference in the disk scale lengths of the stellar mass 
and star formation but it is hard to detect because our proxy for \m\,
is dominated by the youngest stars. At the high mass end, galaxies have 
more dust so star formation could be preferentially obscured at small radii.
Consequently, the \ha\, could appear to be less centrally concentrated
than the star formation is in reality, making the inferred size larger
(see e.g. {Simpson} {et~al.} 2015).
Taken together, these effects could contribute to the trend 
of increasing $r_s(H\alpha)/r_s(H_{F140W})$ with stellar mass. 
However, as described in \S\,7.1, there are a number of other
observational effects that work in the opposite direction, 
decreasing the $r_s(H\alpha)/r_s(H_{F140W})$ at high masses. 
Dust will also obscure the stellar continuum emission, 
meaning that the stellar mass could also be more concentrated than observed. 
Age gradients will also change the M/L ratio, again adding
more stellar mass at the center. 
{Szomoru} {et~al.} (2013) estimate that galaxies are $\sim25\%$ more compact in 
mass than in light.
AGN contributing line emission to the \ha\, profiles will 
also work to decrease this ratio by adding extra flux and decreasing 
the size of the star formation. 
In sum, it seems more likely that observational effects will  
increase the $r_s(H\alpha)/r_s(H_{F140W})$ with mass (and generally) 
than decrease it but as the effects act in both directions we cannot say with 
certainty which are more important. 

While many observational effects could contribute to the the mass dependence 
of the size ratio, this effect may also have a physical explanation.
More massive galaxies have older mean ages. This means that a larger
fraction of their star formation took place at earlier cosmic times.
Hence, it is perhaps then reasonable that their 
stellar mass -- the integral of their past star formation history --
would be more compact than the gas disks with ongoing 
star formation. On the other hand, low mass 
galaxies have younger mean ages, which means their
mass-weighted sizes are closer to the sizes of their star forming disks. 


\subsection{Above and Below the Main Sequence}
\begin{figure*}[hbtf]
\centering
\includegraphics[width=\textwidth]{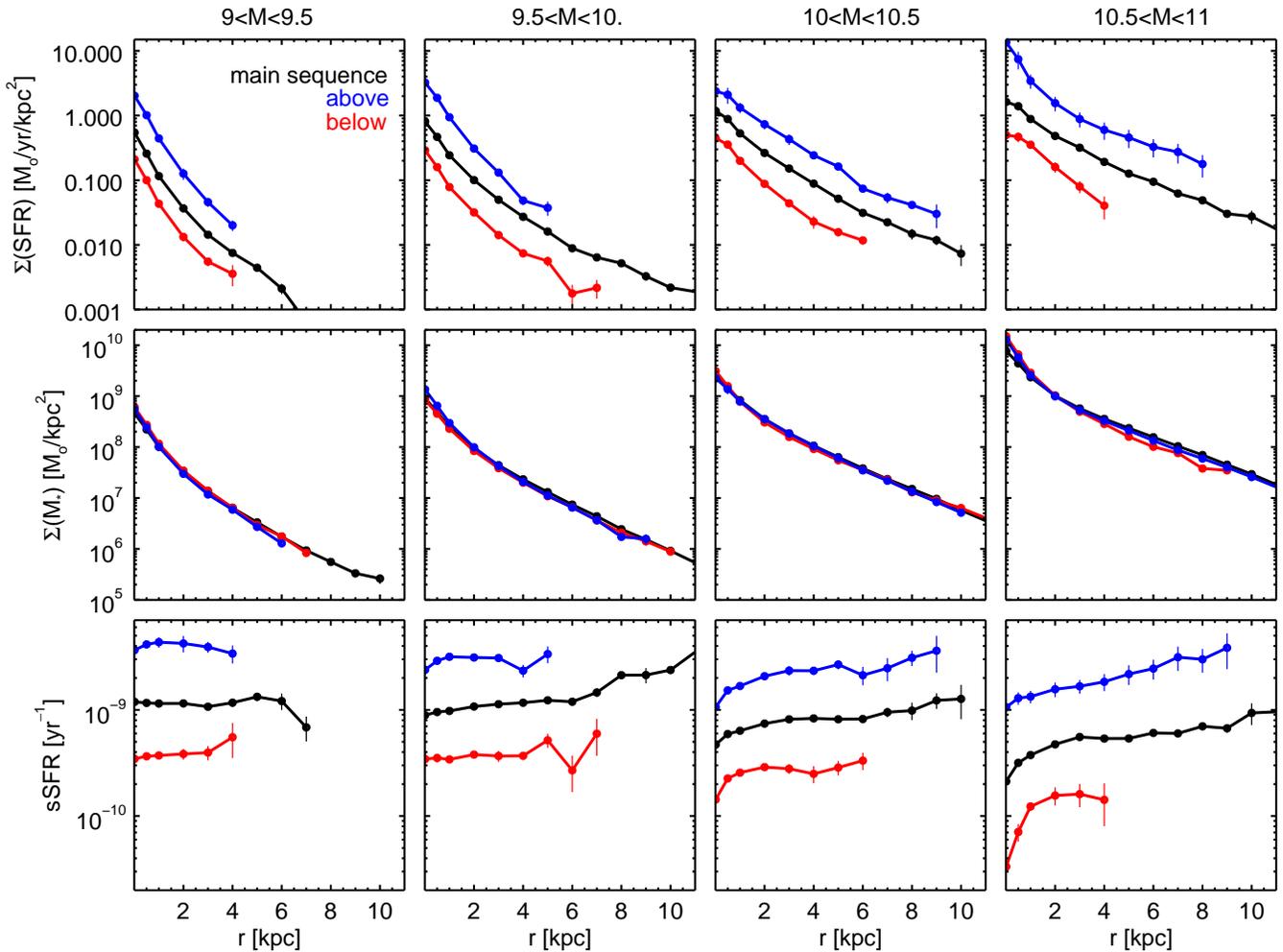}
\caption{Radial surface brightness profiles of SFR, \m, and their ratio 
sSFR as a function of \m\, and SFR. The colors delineate position 
with respect to the star forming `main sequence':
above (blue), on (black), and below (red). 
SFR and \m\ profiles are \ha\ and \f\ profiles scaled to the total SFR(UV+IR)
and \m with all the caveats described in \S8.1. 
Out to distances as great as 8\,kpc from the galactic center, 
star formation is enhanced in galaxies above and depressed in galaxies
below the star forming main sequence. This is also true of the 
specific star formation rate. In general, the radial distribution of \m\ 
is similar on, above, and below the main sequence on average. 
It becomes slightly more centrally concentrated in galaxies above and
below the main sequence at the highest masses, as shown in 
Fig.\,\ref{fig:highMprofs}. There are two take home messages from this figure: 
1. The star formation rate, on average, is always the highest in the 
centers of galaxies. 
2. The radial distribution of star formation depends more strongly
on \m\ than SFR at fixed mass (a galaxy's position with respect to the 
main sequence).
\label{fig:physProfs}}
\end{figure*}

Here we return to the profiles above and below the star-forming main sequence,
that is, for galaxies with
relatively high and relatively low star formation rates for their stellar mass.
{Whitaker} {et~al.} (2012) showed that the
SEDs of galaxies above and below the main sequence are different from those on it. 
Above the main sequence, the SEDs are dusty but blue which they interpreted as indicative of 
AGN or merger-induced starbursts. 
Below the main sequences, the SEDs are not dusty but red, 
which they interpreted as indicative of star formation being shut down. 
Additionally, {Wuyts} {et~al.} (2011a) 
showed that galaxies above and below the main sequence were structurally 
more compact and centrally concentrated than galaxies on the star forming main 
sequence. 

Hints as to what physical processes are driving a galaxy above or below the main 
sequence are given by these trends in stellar structure and SED shape. The next 
key piece of information is \textit{where} the star formation is enhanced above and 
suppressed below the star forming sequence, which we show here.
For instance, if the primary physical processes driving galaxies above the main sequence
are AGN or central starbursts, we would expect \ha\, to be enhanced in 
the center but not at larger radii. If quenching is driven by processes 
acting from the center and progressing from the inside outward, we would
galaxies below the main sequence to have a decrease in \ha\, primarily in 
the center. 

We characterize galaxies with respect to the star formation 
main sequence using their total SFR(IR+UV)s which reflect 
the total obscured+unobscured ionizing flux from young stars.
As described in \S5.2,
we find that above the main sequence, the \ha\ is 
enhanced at all radii; below the above the main sequence the \ha\ is 
depressed at all radii.  
In Fig.\,\ref{fig:physProfs} we show SFR, \m, and sSFR profiles made by 
scaling our \ha\ profiles using the integrated SFR(IR+UV)/SFR(\ha)
and \f\ profiles using the integrated \m$/L_{F140W}$
with all the associated caveats described in \S7.1. 
Because the integrated \m$/L_{F140W}$ decreases with increasing SFR at 
fixed mass, 
the offset in the \f\ light profiles shown in middle panels of Fig.\,\ref{fig:obsprofs} 
disappears in the \m\ profiles shown middle panels of Fig.\,\ref{fig:physProfs}.  
At fixed mass, galaxies are brighter above the main sequence and fainter below 
but the underlying mass profiles are fairly similar at all SFRs
(although see next section for a discussion of the highest masses). 
On the other hand, the dust attenuation increases with increasing SFR at 
fixed mass. Acting in concert, dust and age mean that the \ew\ profiles
shown in the bottom panels of Fig.\,\ref{fig:obsprofs} likely underestimate
the true difference in sSFR above, on, and below the main sequence. 
In the bottom panels of Fig.\,\ref{fig:physProfs} the trends in sSFR are
enhanced after accounting for dust and age.  

The most robust conclusion we can draw about the 
radial distribution of star formation, an inferred quantity, is that 
star formation in the disk between $2-6$\,kpc is enhanced above the main 
sequence and suppressed below the main sequence. 
This, in turn, has several important implications.

First, our results constrain the importance of AGN emission
above the main sequence. One possibility is that galaxies above the
star forming main sequence are there because the bright UV+IR emission 
of an AGN was incorrectly interpreted as star formation. 
In this case, the \ha\ emission would be elevated in the center but
 the same as on the main sequence throughout the rest of the disk.
This, however, is not what we observe: the \ha\, in the disk from 2-6\,kpc is 
elevated, meaning that galaxies are not only above the main sequence 
due to misinterpreted AGN.

Second, because \ha\, is an independent indicator of star formation,
the fact that it is enhanced at all radii confirms that the scatter in 
the main sequence is real and due to variations in the star formation rate at fixed
mass. If the observed main sequence scatter were due exclusively to measurement
errors in the UV+IR SFRs, the \ha\, should not be enhanced or depressed in
concert, but it is. 

Third, the profiles provide information
on the importance of mergers and galaxy encounters
``pushing'' galaxies above the main sequence. It is well established
that interaction-driven gravitational torques can
funnel gas to the center of a galaxy inducing a burst of star formation
 (e.g., {Hernquist} 1989; {Barnes} \& {Hernquist} 1991, 1996; {Mihos} \& {Hernquist} 1996).
However, in idealized merger simulations, {Moreno} {et~al.} (2015) show 
that while star formation is enhanced in the central kpc of interacting galaxies, 
it is \textit{suppressed} everywhere else. This is not what we observe:
the \ha\, in our stacks above the main sequence is enhanced at all radii;
it is not enhanced in the central kpc and suppressed at larger galacto-centric 
radii. Some ambiguity is inherent in the interpretation of an average distribution 
of \ha\, because the distribution of \ha\ in individual galaxies could vary 
significantly from the average. 
Our stacking method cannot distinguish between local enhancements at 
random locations in the disk and global enhancement of the disks of 
individual galaxies. 
Nevertheless, our uniformly higher star formation rates
suggest that major
mergers are not the {\em only} physical process 
driving the elevated star formation in galaxies above the star forming
main sequence.

Below the main sequence, it is possible
the dominant processes suppressing star formation act primarily in the 
centers of galaxies where AGN live, bulges grow, and timescales are short. 
If this were the case, we would expect \ha\, to be lower in the center of the 
galaxies but unchanged at large radii. Again, this is not what we observe:
below the main sequence \ha\, is suppressed at all radii, indicating that the
physical mechanisms suppressing star formation must act over the whole disk, 
not exclusively the center. 

Instead perhaps, for stellar masses below 
$\textrm{M}\sim3\times10^{10}\textrm{M}_\odot$,
some cosmological hydrodynamic simulations ({Sparre} {et~al.} 2015) and 
models ({Dutton}, {van den Bosch}, \&  {Dekel} 2010; {Kelson} 2014) have suggested that a galaxy's 
position in the SFR-\m\ plane is driven by its mass accretion history.
In this schema, galaxies living below the main sequence had early 
formation histories and galaxies above the main sequence had later
formation histories. {Sparre} {et~al.} (2015) show that in Illustris, most of 
the scatter in the star-forming main sequence
is driven by these long scale ($\gtrsim500$Myr)
features of galaxies' formation trajectories rather than short-term 
stochasticity. {Dutton} {et~al.} (2010) predict based on this model for 
main sequence scatter that the size of gas disks should be the same
above and below the main sequence. Consistent with this prediction,  
for masses below $\textrm{M}\sim10^{10.5}\textrm{M}_\odot$, 
we do not see significant differences in \ha\, sizes above and below 
the main sequence, although the error bars are large 
(see Table\,2 for values). 
The fact that 
the average radial distribution of \ha\ does not have wildly different structure
above and below the main sequence perhaps makes more sense in the
context of scatter driven by longer timescale variations 
in the mass accretion history as opposed to some ubiquitous 
physical process.
In other words, the similarity of the radial profiles appears 
consistent with a simple model in which the overall star formation 
rate scales with the gas accretion rate (averaged over some
timescale) and the gas distributes itself in similar structures regardless of 
its accretion rate. 
It will be interesting to compare the observed gas distributions 
directly to those in galaxy formation models. 

Regardless of the physical reasons, across the SFR-\m\, plane two 
important features are consistent. 1) The observed \ha\, distribution is always
centrally peaked. 2) The observed EW(\ha) is never centrally peaked. 

\subsection{Bulge growth and quenching at high masses?}
\begin{figure*}
\includegraphics[width=\textwidth]{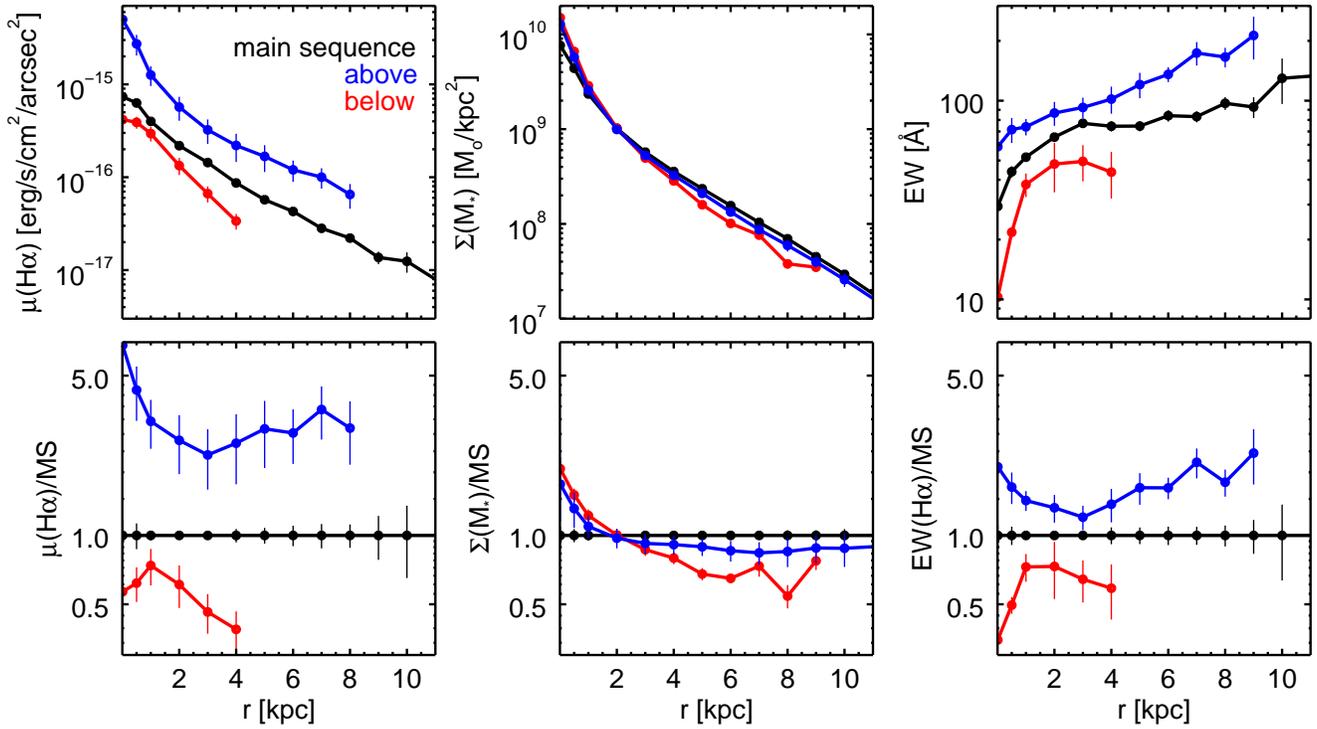}
\caption{Top row, radial profiles of \ha, \m, and \ew\, for galaxies with 
$\textrm{M}=10^{10.5}-10^{11}\textrm{M}_\odot$. 
As previously, the colors delineate position 
with respect to the star forming `main sequence':
above (blue), on (black), and below (red). 
The bottom row shows these profiles normalized by the main 
sequence (with the black line divided out).
Above the main sequence, the \ha\ (and \ew) is enhanced at all radii,
but somewhat more so at small and large radii. 
Below the main sequence, the \ha\ (and \ew) is suppressed at all 
radii, but somewhat more so small and large radii. 
There appears to be excess central stellar 
mass similarly above and below the main sequence.
\label{fig:highMprofs}}
\end{figure*}

\begin{figure}
\centering
\includegraphics[width=0.5\textwidth]{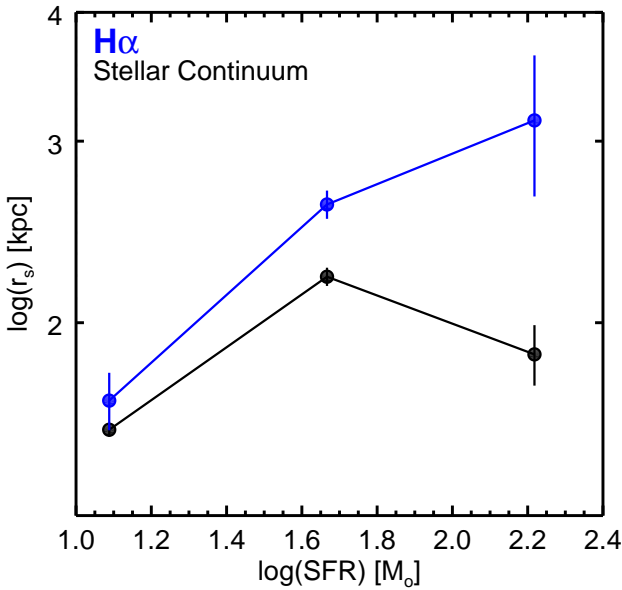}
\caption{Relation between disk scale length and SFR in \ha\, and 
\f\, emission for galaxies with $10.5<{\rm log(M_*)}<11.0$. In \f\, 
emission, the disk scale length is smaller above and below the 
main sequence than on it. In \ha\, emission, the disk scale length
below the main sequence remains smaller than on the main sequence
but is larger above it. 
\label{fig:highMsize}}
\end{figure}

While \ha\ is enhanced at all radii in galaxies above the main sequence 
and suppressed at all radii below the main sequence, in the high mass bin
($\textrm{M}=10^{10.5}-10^{11}\textrm{M}_\odot$), the
trends appear to have some radial dependence as well. 
To examine trends at high masses in more detail, in Fig.\,\ref{fig:highMprofs} 
we show the same the radial profiles of \ha, \f, and 
\ew\ above, on and below the main sequence as in Figs.\,\ref{fig:obsprofs} and 
\ref{fig:physProfs}.
Here we also normalize by the main sequence profiles to highlight differences. 

Above the main sequence, there is a central excess in \ha\ emission
(left panels of Fig.\,\ref{fig:highMprofs}).
The cause of this excess is difficult to interpret:
it could be due to an AGN or extra
star formation in the central regions or both. 
As mentioned in \S\,2.3, galaxies with X-ray luminosity 
$L_x>10^{42.5}{\rm erg\,\,s}^{-1}$ or a 
very obvious broad line component are excluded from the analysis
in this paper. The excess central \ha\ emission exists
even when galaxies hosting obvious 
AGN are excluded. However, with a very conservative
cut on broad line AGN in which galaxies with even marginal elongation
in the spectral direction are excluded, the central excess in \ha\, disappears. 
Hence, it is possible that this central 
enhancement is driven primarily by emission from AGN.
If it is due to an AGN, it could
suggest that supermassive black holes are growing in this 
region of parameter space. If it is due to star formation, it could 
indicate that bulge construction is underway, consistent 
with the growing prominence of bulges observed during this epoch
({Lang} {et~al.} 2014). We note that because
the IR/\ha\, in this bin is so high, it is likely that the excess in 
central ionizing flux (either from star formation or an AGN) would 
actually be even larger if it were not attenuated. 

If the high SFRs in galaxies above the main sequence are fueled by 
elevated gas accretion rates, the disks of these galaxies are likely 
to be gas-rich. In these gas-rich environments, it has been suggested
that gravitational torques 
induced by violent disk instability could drive gas rapidly inward by viscous 
and dynamical friction 
({Noguchi} 1999; {Dekel}, {Sari}, \&  {Ceverino} 2009a; {Krumholz} \& {Burkert} 2010; {Bournaud} {et~al.} 2011; {Genzel} {et~al.} 2011; {Forbes}, {Krumholz}, \&  {Burkert} 2012; {Cacciato}, {Dekel}, \&  {Genel} 2012; {Elmegreen}, {Zhang}, \&  {Hunter} 2012; {Dekel} {et~al.} 2013; {Forbes} {et~al.} 2014). 
Once in the center this gas could fuel the bulge and/or black hole 
growth evidenced by the excess central \ha\ emission. 

During the epoch $0.7<z<1.5$ in this mass range 
($\textrm{M}=10^{10.5}-10^{11}\textrm{M}_\odot$)
the quenched fraction roughly doubles (from $\sim30-60\%$). 
Since the SFRs of galaxies must fall below the main 
sequence on their way to quenchdom, this region of parameter space would
be a good place to look for hints as to how galaxies quench. 
Relative to the main sequence, the \ha\ below the main sequence 
appears to be depressed in the center (Fig.\,\ref{fig:highMprofs} bottom left).  
The \ha\, profile also appears depressed relative to the main sequence
at larger radii, a manifestation of its smaller scale radius (Fig.\,\ref{fig:highMsize}).
That is, we find that below the main sequence, the star-forming disk 
of \ha\ emission is both less centrally concentrated and more compact. 

In addition to the \ha\ in the centers of galaxies below the main sequence 
being depressed relative to galaxies on the main sequence,
it is also depressed relative to the \f\ emission
(Fig.\,\ref{fig:highMprofs}, top right). 
Interpreted as sSFR, this means that the stellar mass doubling time in the 
centers of these galaxies is significantly lower than at larger radii. 
Centrally depressed sSFR has been taken as evidence of inside-out 
quenching ({Tacchella} {et~al.} 2015a). Here we show this for the first time explicitly 
below the main sequence where it is most straight-forward to 
interpret in the context of star formation quenching. 
That being said, it should be noted that although the \ha\, is 
centrally depressed in two interesting
relative senses (relative to the \f\ and relative to the main sequence \ha), 
in an absolute sense,
the \ha\, is \textit{not} centrally depressed, it is centrally peaked. 
That is, on average, there is not a hole in the observed \ha\ emission
at the centers of massive galaxies below the main sequence. 
So while we may be seeing some suppression of star formation in the center
of these galaxies below the main sequence, it is not `quenching' 
in the standard sense of a complete cessation of star 
formation. 

Our findings could be viewed in the context of an evolutionary 
pathway from bulge growth to quenching 
(e.g., {Wuyts} {et~al.} 2011a; {Lang} {et~al.} 2014; {Genzel} {et~al.} 2014b; {Tacchella} {et~al.} 2015a).
Consistent with {Wuyts} {et~al.} (2011a), we find excess central stellar 
continuum emission similarly above and below the star forming 
sequence. {Wuyts} {et~al.} (2011a) suggests that this structural similarity 
could indicate an evolutionary link between the galaxies above 
and below the main sequence. 

AGN can in principle drive gas out of the centers of their 
host galaxies, efficiently removing the fuel for star formation 
(see e.g., {Croton} {et~al.} 2006).
Large bulges are also in principle capable of stabilizing galaxy disks
and suppressing star formation from the inside-out 
(`gravitational quenching' {Martig} {et~al.} 2009; {Genzel} {et~al.} 2014b).
Observationally, it seems that regardless the physical cause,
galaxies quench after reaching a stellar surface density threshold
(e.g. {Franx} {et~al.} 2008). 
Whatever process is underway above
the main sequence, there are theoretical indications that it is capable
of suppressing star formation. Some authors argue this 
occurs from the inside-out. The deep depression in \ew\, in 
the centers of galaxies below the main sequence could be taken as 
evidence for one of these quenching mechanisms acting in this way. 

One remaining mystery, as shown in Fig.\,\ref{fig:highMsize} 
is that the \ha\, disks 
have much smaller sizes below the main sequence than on or above it. 
It is possible that the galaxies below the main sequence formed 
earlier than the galaxies on or above the main sequence at this 
redshift and hence the galaxies above the main sequence 
are not actually direct progenitors of those below.
It is also possible
that these galaxies underwent some sort of compaction
on their way to quenching (e.g.  {Dekel} \& {Burkert} 2014; {Zolotov} {et~al.} 2015)
The most robust thing we can say is that below the main sequence, 
\ha\, seems to be both less centrally concentrated and less extended. 
How exactly this should be interpreted is unclear without the 
aid of simulations.

\subsection{The average spatial distribution of star formation from $z=0.7-1.5$}
\begin{figure*}[hbtf]
\centering 
\includegraphics[width=\textwidth]{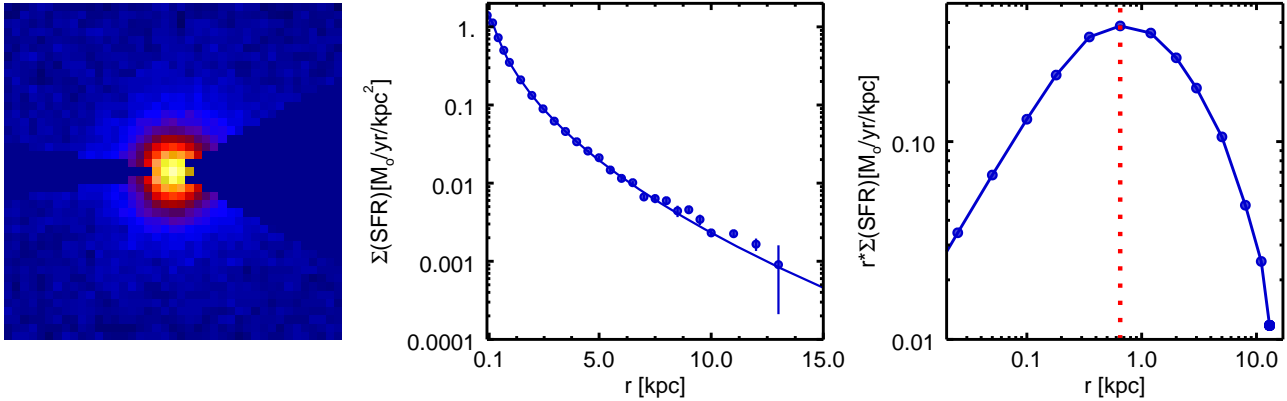}
\caption{Stack of all \ha\, for the redshift epoch $0.7<z<1.5$ and the corresponding
 average radial distribution (left and center.) Weighting by area shows where the average 
 star formed during this epoch: $\sim0.75$\,kpc from the center of it's home galaxy. 
\label{fig:avstar}}
\end{figure*}

In \S\,4\,\&5 we determined the radial profiles of star formation 
as a function of \m\, and SFR. Here we briefly analyze the radial distribution
of all star formation at this epoch, that is, at what distance 
from the center of a galaxy is a star most likely to form.  
The average \ha\, image of all selected galaxies is shown in Fig.\,\ref{fig:avstar}.
This is the average spatial distribution of \ha\, in galaxies during the epoch 
$0.7<z<1.5$. 
Each galaxy has an \ha\, map with a depth of 2 orbits on HST. 
We summed the \ha\, maps of 2676 galaxies, creating the 
equivalent of a 5352 orbit \ha\, image.
This average \ha\, image is deepest \ha\, image in existence 
for galaxies at this epoch.
With this stacked 5352 orbit
HST image, we can trace the radial distribution of \ha\, down to a surface
brightness limit of 
$1\times10^{-18}\,\textrm{erg}\,\textrm{s}^{-1}\,\textrm{cm}^{-2}\,\textrm{arcsec}^{-2}$.
This allows us to map the distribution of 
\ha\, emission out to $\sim 14$\,kpc where the star formation surface density 
is $4\times10^{-4}\,\textrm{M}_\odot\,\textrm{yr}^{-1}\,\textrm{kpc}^{-2}$ 
({Kennicutt} 1998).

Weighting the radial profile of \ha\, by area shows its probability distribution. 
The \ha\, probability distribution has a peak, the expectation value,
at 0.75\,kpc. 
Note, we did not normalize by the \f\, flux here so the expectation value 
reflects the most likely place for a random HII region within a galaxy to exist. 
If we interpret \ha\, as star formation then during the epoch $0.7<z<1.5$, 
when $\sim33\%$ of the total star formation in the history 
of the universe occurred, the most likely place for a new 
star to be born was 0.75\,kpc from the center of its home galaxy.

\section{Conclusions}
In this paper, we studied galaxy growth through star formation during the 
epoch $0.7<z<1.5$ through a new window provided by
the WFC3 G141 grism on HST. 
This slitless grism spectroscopy from space, with its combination
of high spatial resolution and low spectral resolution gives spatially resolved
\ha\ information, for 2676 galaxies over a large swath of the SFR-\m\ plane.
\ha\ can be used as a proxy for star formation, although there are 
many uncertainties (\S\,7.1). 
The most important new observational result of our study is the behavior of the \ha\ 
profiles above and below the main sequence: remarkably, star formation is 
enhanced at all radii above the main sequence, and suppressed at all radii 
below the main sequence (Fig.\,\ref{fig:physProfs}).  
This means that the scatter in the star forming sequence is real.  
It also suggests that the primary mode of star formation is similar 
across all regions of this parameter space.

Across the expanse of the SFR-\m\ plane, the radial distribution of star formation can be characterized in the following way.  
Most of the star formation appears to occur in disks (Fig.\,\ref{fig:msprofs}), 
which are well-aligned with the stellar distribution (Fig.\,\ref{fig:rotStacks}).  
To first order, \ha\ and stellar continuum emission trace each other quite well.  
On average, the \ha\ surface density is always highest in the centers of galaxies, 
just like the stellar mass surface density.  
On the other hand, the \ew, and the inferred specific star formation rate, is, on average, 
{\it never} highest in the centers of galaxies (Fig.\,\ref{fig:obsprofs}).  
Taken at face value, this means that star formation is slightly more extended 
than the existing stars (Fig.\,\ref{fig:mass_size}), 
demonstrating that
galaxies at this epoch are growing in size due to star formation.

The results in this study can be extended in many ways. In principle, the same
dataset can be used to study the spatial distribution of [O\,{\sc iii}] emission
at higher redshifts, although it is more difficult to interpret and
the fact that it is a doublet poses practical difficulties. With submm interferometers
such as NOEMA and ALMA the effects of dust obscuration can be mapped. Although
it will be difficult to match the resolution and sample size that we reach in this study,
this is crucial
as dust is the main uncertainty in the present analysis. Finally, joint studies of the
evolution of the distribution of star formation and the stellar mass can provide constraints
on the importance of mergers and stellar migration in the build-up of present-day disks.


\section{Appendix}
In this paper we investigate the average radial distribution of \ha\, emission by 
stacking the \ha\, maps of individual galaxies and computing the flux in circular 
apertures on this stack. With this methodology, we average over the distribution 
of inclination angles, position angles, and sizes of galaxies that go into each stack.
The simplicity of this method has a number of advantages. First, it requires no 
assumptions about the intrinsic properties of galaxies. Second, it allows us to measure
the average size of the \ha\, distribution in the star forming disk. Finally, because
the image plane is left in tact, we can correct for the PSF. 

To complement this analysis, here we present the average deprojected, 
radially-normalized distribution of \ha. We do this to test the effect of projection and 
a heterogenous mix of sizes on the shape of the radial profile of \ha, to ensure
trends were not washed out with the simpler methodology employed in the rest 
of the paper.

To do this, we use GALFIT ({Peng} {et~al.} 2002) to derive the effective 
radius, axis ratio, and position angle of each galaxy from its \f\ stellar continuum image. 
We correct for the inclination angle of each galaxy by 
deprojecting the (x,y) pixel grid of it's image based on the inclination angle
implied by the axis ratio. The surface brightness profile is computed by 
measuring the flux in deprojected radial apertures. 
In practice, this is done simply by extracting the radial profile of each galaxy in elliptical
apertures defined by the position angle, axis ratio, and center of the \f\, image. 
The extraction apertures were normalized by the \f\, effective radius of each galaxy. 
A radial profile in deprojected, $r_e$-normalized space is derived
for each galaxy. These individual galaxy profiles are flux-normalized by their 
integrated \f\, magnitude and summed to derive the mean radial distribution. 

\begin{figure*}[hbtf]
\centering
\includegraphics[width=\textwidth]{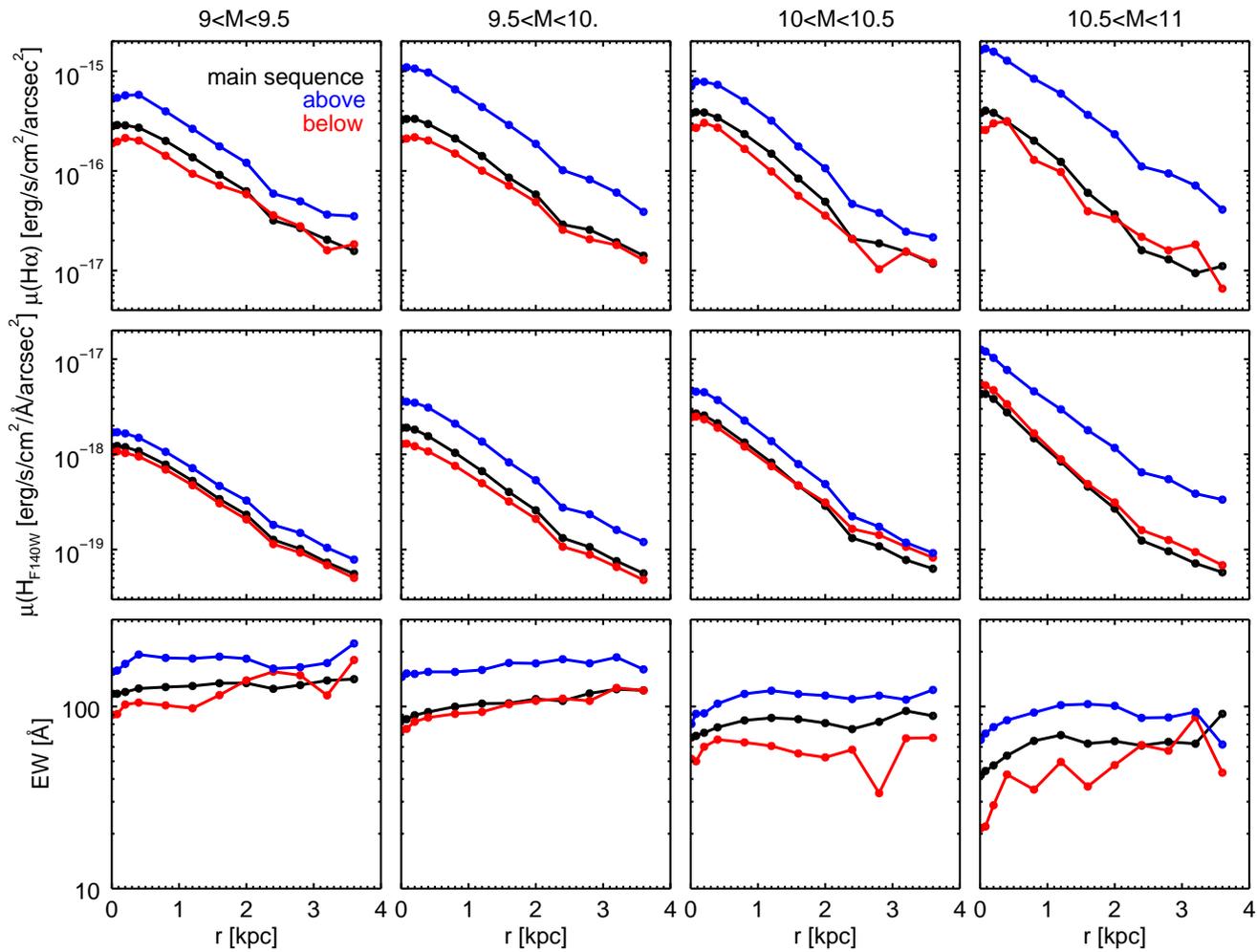}
\caption{Average de-projected, $r_e$-normalized radial profiles of \ha, \f, and \ew\, 
as a function of mass. 
\label{fig:NFSprofs}}
\end{figure*}

The average de-projected, $r_e$-normalized radial profiles of \ha, \f, and \ew\, 
are shown in Fig.\,\ref{fig:NFSprofs}. 
In general, the qualitative trends seen here are the same as those
described in the main text. 
For the region $0.5<r_e<3$ the radial profile of \ha\ remains consistent with an 
exponential all masses, above, on, and below the star forming sequence. 
The radial profiles of both \ha\ and \f\ are somewhat 
less centrally peaked than the analogous profiles in Fig.\,\ref{fig:obsprofs}. 
This is expected of disk-dominated galaxies under different orientation angles as
flux from the disk of edge-on galaxies could be projected onto the center. 
Additionally, stacking galaxies of different sizes can result in a somewhat steeper
(higher n) profile than the individual galaxies that went into it (see  {van Dokkum} {et~al.} 2010).
Because the shapes of the \ha\, and  \f\, profiles are similarly effected by 
deriving the profiles with this different methodology, the shape of the \ew\ 
profiles remain largely unchanged.

\end{document}